\documentclass[
reprint,
 amsmath,
 amssymb,
 aps,
 pra,
]{revtex4-2}

\usepackage{graphicx}
\usepackage{dcolumn}
\usepackage{bm}
\usepackage[hidelinks]{hyperref}

\usepackage{xcolor}
\usepackage{overpic}
\usepackage{comment}
\usepackage{dsfont}
\usepackage{physics}

\usepackage{cleveref}
\crefname{equation}{Eq.}{Eqs.}
\Crefname{equation}{Equation}{Equations}
\crefname{figure}{Fig.}{Figs.}
\Crefname{figure}{Figure}{Figures}
\crefname{section}{Sec.}{Secs.}
\Crefname{section}{Section}{Sections}
\crefname{appendix}{Appendix}{Appendices}
\Crefname{appendix}{Appendix}{Appendices}

\renewcommand{\d}{\mathrm{d}}
\newcommand{\ee}{\mathrm{e}}
\newcommand{\ii}{\mathrm{i}}

\newcommand{\crefs}[1]{Figs.~\ref{#1}}
\newcommand{\Crefs}[1]{Figures~\ref{#1}}

\renewcommand{\Tr}[1]{\mathrm{Tr}\left[#1\right]}
\begin{document}
\title{Quantum synchronization blockade induced by nonreciprocal coupling}

\author{Tobias Kehrer}
\author{Christoph Bruder}
\affiliation{Department of Physics, University of Basel, Klingelbergstrasse 82, CH-4056 Basel, Switzerland}

\date{\today}

\begin{abstract}
Recently, the synchronization of coupled quantum oscillators has attracted a great deal of interest.
Synchronization requires driven constituents, and in such systems, the coupling can be designed to be nonreciprocal. 
Nonreciprocally coupled oscillators exhibit a rich variety of behavior including active traveling-wave-type states.
In this work, we study the interplay of three competing synchronization mechanisms in a setup of two nonreciprocally coupled quantum van der Pol oscillators.
One of the oscillators is driven externally which induces phase locking.
A dissipative interaction leads to antiphase locking, whereas a coherent interaction nurtures bistable phase locking and active states.
We approximate the phase diagram of the quantum case by evaluating the synchronization measure of a perturbation expansion of the steady state.
Effective unidirectional interactions lead to synchronization blockades between the undriven oscillator and the external drive as well as between both oscillators.
Furthermore, we study the phase diagrams of two and three oscillators in the mean-field limit and find highly nontrivial active states.
\end{abstract}

\maketitle

\noindent

\section{Introduction}
In the last decade, research on quantum synchronization has attracted a great deal of attention.
One of its goals is the study of quantum analogues of synchronization in classical dynamics~\cite{Synch_Pikovsky,Synch_Strogatz,RevModPhys.77.137,Balanov2008}, i.e., the entrainment of oscillation frequencies or phases of oscillators among each other or to external signals.
The building blocks of classical synchronization are limit-cycle oscillators that are characterized by closed trajectories in phase space stabilized by gain and damping.
Quantum analogues of limit cycles have been considered in a number of systems like van der Pol oscillators~\cite{PhysRevLett.111.073603,Synch_vdP_Lee,Synch_vdP_Walter,Chia2020,PhysRevResearch.3.013130} as well as few-level spin-like objects~\cite{PhysRevLett.121.053601,PhysRevLett.121.063601,PhysRevA.101.062104}.
Some of the predicted quantum synchronization effects have no classical 
counterpart. This includes effects that are based on 
quantum features like entanglement~\cite{PhysRevLett.121.063601,PhysRevA.91.012301,PhysRevLett.111.103605,manzano2013synchronization}. Another example is 
the synchronization blockade, the destructive interference of coherences which can lead to the suppression of synchronization~\cite{PhysRevLett.118.243602,PhysRevA.108.022216,PhysRevA.110.042203}.

Interactions between two agents $A$ and $B$ are called nonreciprocal if the response of $A$ to an action of $B$ differs from the response of $B$ to an action of $A$.
Nonreciprocal interactions can only appear in nonequilibrium systems~\cite{PhysRevX.5.011035}, in particular, in active matter, i.e., systems composed of active agents~\cite{Ramaswamy2010,Schweitzer2019}, and have been intensively studied in classical models. 
Prime examples of such active states are the so-called traveling-wave states.
In nonreciprocal models like the Lotka-Volterra predator-prey model~\cite{Lotka1920,VOLTERRA1926,Bacaer2011} these states are associated to two different agents one of which (predator) is hunting the other (prey).
More recently, phase transitions~\cite{Fruchart2021} and frustration~\cite{PhysRevX.14.011029} in systems of nonreciprocal oscillators have been investigated.
First steps towards nonreciprocity in quantum systems have been taken, e.g., in non-Hermitian quantum mechanics~\cite{PhysRevLett.77.570}, cascaded networks~\cite{PhysRevA.94.043841,Lorenzo2022}, and topological networks~\cite{Wanjura2020,PhysRevResearch.5.023021}.
Lately, investigations of the effects of nonreciprocal interactions on quantum synchronization have started~\cite{PhysRevX.15.011010}. 

\begin{figure}[t]
    \centering
    \includegraphics[width=8.6cm]{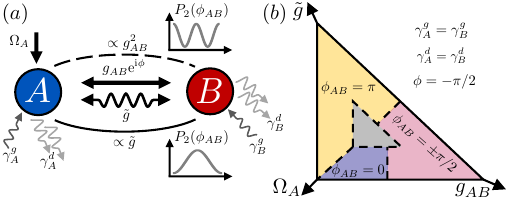}
    \caption{Schematic overview of two coherently and dissipatively coupled, driven oscillators.
    (a) Each of the oscillators $A$ and $B$ is subject to single-phonon gain and two-phonon loss.
    The coherent coupling $g^{}_{AB}\ee^{\ii\phi}$ is denoted by a solid double arrow and the dissipative coupling $\tilde{g}$ by a wavy double arrow.
    An external drive $\Omega_A$ represented by a solid arrow is applied to $A$.
    The solid (dashed) arc visualizes (bistable) locking between the oscillators. 
    The insets are qualitative plots of the combined synchronization measure $P_2$, the probability distribution of the relative phase $\phi_{AB}$.
    (b) Schematic regions labeled by the steady-state values of $\phi_{AB}$ at which $P_2$  exhibits a maximum.
    Each corner/arrow head corresponds to the regime in which this parameter is large compared to the others.
    Dashed lines indicate approximate transitions.}
    \label{fig:schematics}
\end{figure}

In this work, we consider systems of two coupled quantum limit-cycle oscillators and study the interplay of three competing quantum synchronization mechanisms: phase locking, antiphase locking, and bistable locking.
These three effects are induced by an external coherent drive that acts on one of the two quantum oscillators as well as by a coherent and dissipative coupling that yield an effective nonreciprocal interaction between the oscillators.
The two couplings can be tuned such that the nonreciprocal interaction even becomes unidirectional~\cite{PhysRevX.5.021025}.
A schematic overview of the phase-locking regimes is presented in \cref{fig:schematics}.
To quantify quantum synchronization, we employ a common measure.
We show that the effective interaction leads to synchronization blockades.
One blockade occurs between the undriven oscillator and the external drive in the unidirectional case when oscillator $A$ does not influence oscillator $B$. 
The second blockade occurs between both oscillators when the effective interaction is close to being unidirectional.
A mean-field analysis reproduces this behavior.
To understand this blockade in the quantum case, we make use of the quantum synchronization measure evaluated for a perturbation expansion of the steady state.

The paper is structured as follows.
In \cref{sec:model}, we introduce the Lindblad master equation which describes the gain and damping processes that stabilize the quantum limit cycles and define suitable quantum synchronization measures.
We start our analysis by considering two coherently coupled oscillators of which one is driven externally in \cref{sec:coherent}.
Then, we introduce a dissipative coupling and study its effect both in the absence and in the presence of the external drive in \cref{sec:dissipative}.
In \cref{sec:blockades}, we analyze the blockades induced by the nonreciprocal interactions.
In the last section, \cref{sec:MF}, we compare the phase diagram of our quantum model to the ones of classical analogues that are defined by the corresponding mean-field equations.

\section{Model}\label{sec:model}
We consider two limit-cycle oscillators stabilized by single-phonon gain at rate $\gamma^g_j$ and two-phonon damping at rate $\gamma^d_j$~\cite{Synch_vdP_Lee},
\begin{align}
    \dot{\rho} =& \mathcal{L}(\rho) = -\ii[H, \rho] + \tilde{\mathcal{L}}(\rho)\,,\label{eq:masterEQ1}\\
    \tilde{\mathcal{L}}(\rho) =& \frac{\gamma^g_A}{2}\mathcal{D}[a^\dagger](\rho) + \frac{\gamma^g_B}{2}\mathcal{D}[b^\dagger](\rho)\nonumber\\
    &+ \frac{\gamma^d_A}{2}\mathcal{D}[a^2](\rho) + \frac{\gamma^d_B}{2}\mathcal{D}[b^2](\rho)\,.\label{eq:Leom}
\end{align}
The operators $a^{(\dagger)}=a^{(\dagger)}_A$ and $b^{(\dagger)}=a^{(\dagger)}_B$ denote the annihilation (creation) operators of system $A$ and $B$.
The Hamiltonian $H$ will be defined in the individual sections below and contains coherent drive and coupling terms.
Later, we will introduce an additional dissipative coupling between both oscillators to create an effective unidirectional coupling.
A schematic overview of the system is given in \cref{fig:schematics}.

To study quantum synchronization phenomena in this model, we have to choose an appropriate quantitative measure of synchronization. 
In previous works, several measures have been defined~\cite{PhysRevLett.111.073603,phase_dist_Hush,PhysRevA.91.012301,Weiss_2016,PhysRevLett.121.053601,Jaseem_2020}.
In the present study, we will follow~\cite{phase_dist_Hush,Weiss_2016} and consider probability distributions of phases of quantum oscillators.
These distributions are based on the phase states \cite{phase_dist_Barak}
\begin{align}
    \ket{\phi} &= \frac{1}{\sqrt{2\pi}}\sum_{n=0}^\infty \ee^{\ii n \phi}\ket{n}\,.
\end{align}
For a single oscillator, the measure $P_1$ is given by
\begin{align}
    P_1(\phi) &= \bra{\phi}\rho\ket{\phi} - \frac{1}{2\pi}\nonumber\\
    &= \frac{1}{2\pi}\sum_{n,m=0}^\infty \ee^{\ii (m-n) \phi} \rho_{n,m} - \frac{1}{2\pi}\,,
\end{align}
where $\rho_{n,m}=\bra{n}\rho\ket{m}$.
Similar to the quantum synchronization measure of spins considered in~\cite{PhysRevA.110.042203}, $P_1$ can be rewritten in terms of expectation values of an operator
\begin{align}
    p(\phi) &= \frac{1}{2\pi} \left(\mathds{1} + \sum_{k=1}^\infty (\ee^{-\ii k \phi}\tilde{a}^k + \mathrm{H.c.})\right)\,.\label{eq:popdef}
\end{align}
Here, the operator
\begin{align}
    \tilde{a} &= \sum_{n=0}^\infty \dyad{n}{n+1},~~~
    \tilde{a}^k = \sum_{n=0}^\infty \dyad{n}{n+k}\,,\label{eq:atilde}
\end{align}
resembles the Susskind-Glogower operator~\cite{PhysicsPhysiqueFizika.1.49}.
This leads to
\begin{align}
    P_1(\phi) &= \Tr{p(\phi)\rho} - \frac{1}{2\pi} =  \langle p(\phi)\rangle - \frac{1}{2\pi}\nonumber\\
    &= \frac{1}{2\pi}\sum_{k=1}^\infty \ee^{-\ii k \phi}\langle\tilde{a}^k\rangle + \mathrm{H.c.}\,.\label{eq:P1viaOp}
\end{align}
For a system containing $N$ quantum oscillators, we consider the following synchronization measure
\begin{align}
    P_N(\vec{\phi}\,) &= \bra{\vec{\phi}}\rho\ket{\vec{\phi}} - \frac{1}{(2\pi)^N} = \left\langle\bigotimes_{j=1}^N p(\phi_j)\right\rangle - \frac{1}{(2\pi)^N}\,,\label{eq:PnviaOp}
\end{align}
that is based on tensor products of phase states
\begin{align}
    \ket{\vec{\phi}} &= \bigotimes_{j=1}^N \ket{\phi_j}\,.
\end{align}
In \cref{eq:PnviaOp}, this measure is rewritten as tensor products of $p(\phi_j)$ defined in \cref{eq:popdef}.
Therefore, its terms contain various combinations of $\ee^{-\ii k \phi_j}\tilde{a}^k_j$ that act on the $j$th oscillator and Hermitian conjugates thereof.
Thus, the moments of the phase distributions $P_N$ are given by expectation values of products of $\tilde{a}^{(\dagger) k_j}_j$.

The phase distribution measure $P_2(\phi_{AB})$ of the relative phase $\phi_{AB}=\phi_A-\phi_B$ of two oscillators reads
\begin{align}
    P_2(\phi_{AB}) &= \int\limits_0^{2\pi}\d\phi_B\,P_2(\phi_{AB}+\phi_B, \phi_B) \nonumber\\
    &= \frac{1}{2\pi}\sum_{k=1}^\infty \ee^{-\ii k \phi_{AB}}\langle(\tilde{a}^{\phantom{\dagger}}_A\tilde{a}^\dagger_B)^k\rangle + \mathrm{H.c.} \,. \label{eq:P2viaOp}
\end{align}
Due to the operator structure of $P_N$ mentioned above, we can define the moments of these phase distributions for individual phases $\phi_j$ and relative phases $\phi_{ij}$ as 
\begin{align}
    m^{(n)}_j &= \langle \tilde{a}_j^n \rangle\,,\\
    m^{(n)}_{ij} &= \langle (\tilde{a}^{\phantom{\dagger}}_i \tilde{a}^{\dagger}_j)^n \rangle\,.
\end{align}

\begin{figure}[t]
    \centering
    \includegraphics[width=8.6cm]{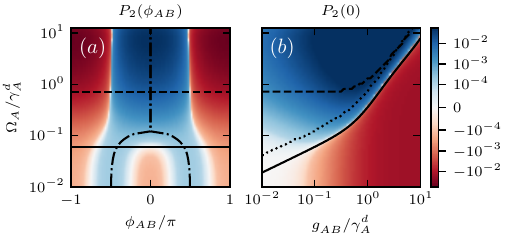}
    \caption{Probability distribution of the relative phase of two oscillators for $\phi=-\pi/2$ and $\tilde{g}=0$.
    (a) Fixed interaction strength $g^{}_{AB}=0.1\gamma^d_A$.
    The dash-dotted black curve denotes the maxima of $P_2$.
    (b) $P_2(0)$ as a function of $g^{}_{AB}$.
    The dotted black curve denotes the transition from two maxima to one maximum.
    In both panels, the color is scaled linear in the interval $[-10^{-4},10^{-4}]$ and logarithmic elsewhere.
    The dashed black curves indicate where $m^{(2)}_{AB}=0$ and the solid black curves indicate where $|m^{(1)}_{AB}|=|m^{(2)}_{AB}|$.}
    \label{fig:P2_Om_vs_phiAB}
\end{figure}

\section{Coherently Coupled Oscillators}\label{sec:coherent}
In previous work~\cite{Synch_vdP_Lee}, two distinct cases have been studied: (i) a single driven limit-cycle oscillator and (ii) two coherently coupled identical limit-cycle oscillators, i.e, with gain and damping rates $\gamma^g_A=\gamma^g_B$ and $\gamma^d_A=\gamma^d_B$ but $\gamma^g_A \neq \gamma^d_A$.
The single oscillator locks to the phase of the external drive with a phase shift of $-\pi/2$.
Note that in the context of quantum synchronization the existence of a single maximum of the synchronization measure at $\phi_0$ is referred to as ``phase locking to $\phi_0$'', i.e., this maximum does not need to be infinitely sharp.
The two coherently coupled oscillators were found to be in the quantum synchronization blockade and exhibit \textit{bistable} phase locking.

Here, we first consider the combination of both cases, i.e., two coherently coupled identical limit-cycle oscillators of which one is driven externally.
In this and the next section, all gain and damping rates are set to be equal $\gamma^d_A=\gamma^d_B=\gamma^g_A=\gamma^g_B$. 
For this choice, the oscillators are in the blockade and are neither in the classical limit $\gamma^d_j \ll \gamma^g_j$ nor in the quantum limit $\gamma^d_j \gg \gamma^g_j$.
The system is described by \cref{eq:masterEQ1} and the Hamiltonian
\begin{align}
    H = \frac{\Omega_A}{2}a^\dagger + \frac{g^{}_{AB}}{2}\ee^{\ii\phi} a^\dagger b  + \mathrm{H.c.}\,.\label{eq:Ham1}
\end{align}
In the original description of the synchronization behavior of identical quantum limit-cycle oscillators~\cite{Synch_vdP_Lee} one can identify two separate locking mechanisms.
First, a driven oscillator $A$ tends to align its phase to the one of the external drive plus a shift of $-\pi/2$.
In the limit where another coupled oscillator $B$ identifies the driven oscillator $A$ as an effective drive, the relative phase between both oscillators will be $\phi_{AB}=\phi_A-\phi_B=\phi+\pi/2$.
The parameter $\phi$ is the complex phase of the coherent coupling between $A$ and $B$, defined in \cref{eq:Ham1}.
Second, the probability distribution of the relative phase for two coherently coupled \textit{undriven} oscillators will exhibit two maxima at different values $\phi_{AB}=\phi, \phi+\pi $.
Therefore, these two locking mechanisms compete in the following sense: depending on the ratio of drive strength and coupling strength, the combined synchronization measure either exhibits one maximum or two maxima.

In \cref{fig:P2_Om_vs_phiAB}, the transition from two maxima to one maximum of the combined synchronization measure for $\phi=-\pi/2$ is visualized.
For small drive strengths, $P_2$ exhibits two maxima at $\phi_{AB}=\pm\pi/2$ that merge into a single maximum at $\phi_{AB}=0$ for sufficiently large drive strength.
In \cref{fig:P2_Om_vs_phiAB}(a) the dash-dotted black curve highlights local maxima of $P_2$ whereas in \cref{fig:P2_Om_vs_phiAB}(b) the dotted curve indicates the point of transition from two maxima to one maximum. 
The maxima merge at values of $\Omega_A$ between the dashed black line where $m^{(2)}_{AB}=0$ and the solid black line that indicates $|m^{(1)}_{AB}|=|m^{(2)}_{AB}|$.

\section{Coherently and Dissipatively Coupled Oscillators}\label{sec:dissipative}
We now add a dissipative coupling $\tilde{g}\,\mathcal{D}[a+b](\rho)$ between the two oscillators to the Lindblad master equation \cref{eq:masterEQ1}.
In this modified setup, the Heisenberg equations of motion of $a$ and $b$ can exhibit an effective unidirectional coupling~\cite{PhysRevX.5.021025}, see \cref{sec:effectivemodel}. This coupling depends on the two possible directions \begin{align}
    g^\text{eff}_{A\rightharpoonup B} &= -\ii g^{}_{AB}\ee^{-\ii\phi} - \tilde{g}\,,\label{eq:geffAB}\\
    g^\text{eff}_{A\leftharpoondown B} &= -\ii g^{}_{AB}\ee^{\ii\phi} - \tilde{g}\,.\label{eq:geffBA}
\end{align}
The influence of oscillator $B$ on $A$ ($A$ on $B$) vanishes for $g^{}_{AB}=\tilde{g}$ and $\phi=(-)\pi/2$, i.e., the effective coupling becomes unidirectional.

\begin{figure}[t]
    \centering
    \includegraphics[width=8.6cm]{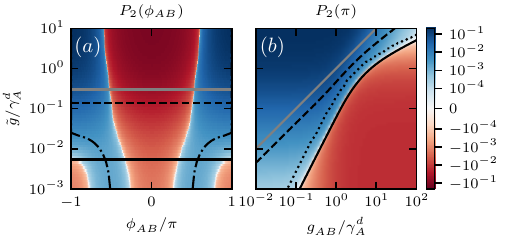}
    \caption{Combined synchronization measure $P_2$ for $\phi=-\pi/2$ and $\Omega_A=0$. (a) Fixed interaction strength $g^{}_{AB}=0.3\gamma^d_A$.
    The dash-dotted black curves denote the maxima of $P_2$.
    (b) $P_2(\pi)$ as a function of $g^{}_{AB}$.
    Here, the dotted curve denotes the transition from two maxima to one maximum of $P_2$. 
    In both panels, the solid gray lines denote $\tilde{g}=g^{}_{AB}$. 
    The dashed black curves indicate where $m^{(2)}_{AB}=0$ and the solid black curves indicate where $|m^{(1)}_{AB}|=|m^{(2)}_{AB}|$.
    The color is scaled linear in the interval $[-10^{-4},10^{-4}]$ and logarithmic elsewhere.}
    \label{fig:P2_vs_g}
\end{figure}

\subsection{No external drive}\label{sec:no_drive}
In~\cite{Synch_vdP_Walter}, it has been shown that two dissipatively coupled quantum limit-cycle oscillators lock to a relative phase $\phi_{AB}=\pi$.
This synchronization behavior is different to the one induced by a coherent coupling with complex phase $\phi=-\pi/2$, see \cref{sec:coherent}.
In \cref{fig:P2_vs_g}(a), we present the combined synchronization measure for a fixed coherent coupling strength, whereas in \cref{fig:P2_vs_g}(b), we vary both the coherent and dissipative coupling strengths to study the transition between both locking mechanisms at $\Omega_A=0$.
For increasing $g^{}_{AB}$ at fixed $\tilde{g}$, four changes occur that are shown in \cref{fig:P2_vs_g}(b): the effective coupling becomes unidirectional (solid gray line), the second moment vanishes $m^{(2)}_{AB}=0$ (dashed black curve), the two maxima of the combined synchronization measure originally at $\phi_{AB}=\pm\pi/2$ turn into a single maximum at $\phi_{AB}=\pi$ (dotted black curve), and the first and second moment become equal $|m^{(1)}_{AB}|=|m^{(2)}_{AB}|$ (solid black curve).
The second moment does not vanish when the effective coupling becomes unidirectional; this feature will be studied in more detail in \cref{sec:blockades}.
For small $g^{}_{AB}$, we recognize that the boundary between one and two locking phases follows the scaling $\tilde{g}\propto g^2_{AB}/\gamma^d_A$ and this behavior is reproduced by the mean-field approximation presented in \cref{sec:MF_pert}.

In the configuration of vanishing drive strength, the system exhibits several symmetries:
first, a global $\mathrm{U}(1)$ symmetry, i.e., the invariance of the Liouvillian $\mathcal{L}$ under the transformation $a_j \to \ee^{\ii\theta}a_j$.
The interaction term $a^\dagger b \to \ee^{-\ii\theta}a^\dagger \ee^{\ii\theta}b = a^\dagger b$ as well as the Lindblad dissipators $\mathcal{D}[L] \to \mathcal{D}[\ee^{\ii k\theta}L] = |\ee^{\ii k\theta}|\mathcal{D}[L] = \mathcal{D}[L]$ are independently invariant under this transformation.
Second, for $\phi=0,\pi$, the Liouvillian is invariant under the transformation $a \to \ee^{\ii\phi}b, b \to \ee^{-\ii\phi}a$.
Here, $\ee^{\ii\phi}a^\dagger b +\ee^{-\ii\phi}a b^\dagger \to \ee^{\ii\phi}\ee^{-\ii\phi}b^\dagger \ee^{-\ii\phi}a +\ee^{-\ii\phi}\ee^{\ii\phi}a^\dagger \ee^{\ii\phi} b = \ee^{\ii\phi}a^\dagger b +\ee^{-\ii\phi}a b^\dagger$ as well as $\mathcal{D}[L]$ are invariant.
Note that $a+b \to \ee^{\ii\phi}(a+\ee^{-2\ii\phi}b)=\ee^{\ii\phi}(a+b)$ for $\phi=0,\pi$.
Third, for $\phi=\pm\pi/2$, the Liouvillian is real $\mathcal{L}=\mathcal{L}^*$ which implies that the steady state $\rho_0=\rho^*_0$ is also real.
Following~\cite{Fruchart2021,PhysRevX.15.011010}, this invariance can be interpreted as a generalized $\mathcal{PT}$ symmetry.
In our setup, this symmetry is defined as the invariance under the consecutive transformations $a \leftrightarrow b$ and $g^{}_{AB} \to -g^{}_{AB}$.
In other words, if the oscillators are exchanged, we arrive again at the same physics if also the sign of $g^{}_{AB}$ is flipped.

\subsection{With external drive}\label{sec:with_drive}
In this section we consider all three parameters $\Omega_A$, $g^{}_{AB}$, and $\tilde{g}$ to be nonzero. There are three competing synchronization effects: First, as described in \cref{sec:coherent}, the external drive defines a preferred phase to which oscillator $A$ locks with a phase shift of $-\pi/2$.
If the coherent coupling with complex phase $\phi$ is small compared to the drive, it leads to a locking of oscillator $B$ such that the relative phase results in $\phi_{AB} = \phi+\pi/2$.
Second, the coherent coupling itself leads to a bistable locking of the relative phase to $\phi_{AB}= \phi, \phi+\pi$.
Third, the dissipative coupling induces locking to $\phi_{AB}=\pi$.

\begin{figure}[t]
    \centering
    \includegraphics[width=8.6cm]{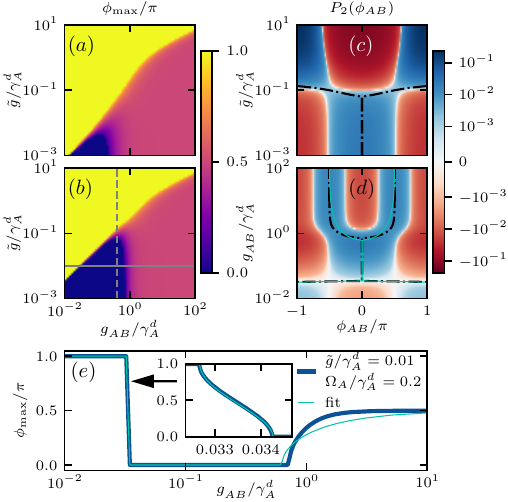}
    \caption{Visualization of different regimes of phase locking.
    The maxima of the combined synchronization measure $P_2$ are located at $\pm\phi_\text{max}$.
    (a) $\Omega_A=0.2\gamma^d_A$.
    (b) $\Omega_A=0.5\gamma^d_A$.
    The dashed gray line indicates the line cut at $g^{}_{AB}=10^{-0.4}\gamma^d_A$ shown in (c) and the solid gray line indicates the line cut at $\tilde{g}=0.01\gamma^d_A$ shown in (d).
    (c), (d) Combined synchronization measures along line cuts highlighted in (b). 
    The dash-dotted black curves highlight the maxima of $P_2$.
    The thin light blue curve corresponds to a fit of the maximum of \cref{eq:Pm} to data of (d).
    (e) The thick blue curve corresponds to the line cut in (b) indicated by the solid gray line, i.e., the dash-dotted black curves in (d).
    The inset shows a zoom to the step-like change of $\phi_\text{max}$.}
    \label{fig:phi_max}
\end{figure}

Two cuts through the three-dimensional phase diagram at $\Omega_A=0.2\gamma^d_A$ and $\Omega_A=0.5\gamma^d_A$ are presented in \crefs{fig:phi_max}(a) and (b).
Three regions of the maxima $\pm\phi_\text{max}$ of $P_2$ can be identified:
first, the bottom left corner corresponds to a dominant drive where the ratio between $\tilde{g}$ and $g^{}_{AB}$ determines the relative phase $\phi_{AB}$, i.e., $0$ or $\pi$ as explained in the beginning of this section and visualized in \cref{fig:schematics}.
Second, in the bottom right corner, in which the coherent coupling $g^{}_{AB}$ dominates, the combined synchronization measure experiences maxima at $\phi_{AB}=\pm\pi/2$.
Third, in the top left corner, where the dissipative coupling dominates, the relative phase reaches $\phi_{AB}=\pi$.
\Crefs{fig:phi_max}(c) and (d) show the combined synchronization measure along two line cuts in \cref{fig:phi_max}(b) where \cref{fig:phi_max}(c) corresponds to the dashed gray line and \cref{fig:phi_max}(d) corresponds to the solid gray line.
In \cref{fig:phi_max}(e) we present the line cut shown in \cref{fig:phi_max}(d) as well as a fit of a model for $\phi_\text{max}$.
This model is defined as the maximum of $P_m$
\begin{align}
    P_m(\phi_{AB}) =& \left(u_1 \frac{g^{}_{AB} \Omega_A^2}{\gamma^3} - u_3 \frac{\tilde{g}}{\gamma}\right)\cos(\phi_{AB}) \nonumber\\
    &+ \frac{u_2 \tilde{g}^2 - u_4 g^2_{AB}}{\gamma^2}\cos(2\phi_{AB})\,,\label{eq:Pm}
\end{align}
with $u_i>0$. 
The parameter $u_1$ ($u_3$) corresponds to a maximum at $0$ ($\pi$) and the parameter $u_2$ ($u_4$) corresponds to maxima at $0$ and $\pi$ (at $\pm\pi/2$).
The powers of the parameters in $P_m$ were obtained by a perturbation expansion of the steady state in the parameters $g^{}_{AB}$, $\tilde{g}$, and $\Omega_A$ with respect to the equal gain and damping rates $\gamma^d_j=\gamma^g_j=\gamma$.
For each of the two cosine terms in \cref{eq:Pm}, we only consider the leading order of each parameter up to a combined third order.
Since in this calculation we truncate the Fock space at a finite occupation number, the values of $u_j$ cannot be obtained.
To get a rough estimate of these values, we fit the maximum of $P_m$ to \cref{fig:phi_max}(e) at $\tilde{g}/\gamma^d_A=0.01$.
The fit $(u_1,u_2,u_3)\approx(11,6.0,8.8)u_4$ shows a good match with the numerical data for $g^{}_{AB}\ll \gamma^d_A$.
Note that this simple model is only suitable for small $\tilde{g}$ and $g^{}_{AB}$. 
For large $g^{}_{AB}/\gamma^d_A$, the transition of $\phi_\text{max}$ from $0$ to $\pi/2$ is captured qualitatively.
The linear dependence $\tilde{g}\propto g^{}_{AB}$ for which the second moment in \cref{eq:Pm} vanishes, see the dashed black curve in \cref{fig:P2_vs_g}(b), appears to be valid even slightly above $g^{}_{AB}=\gamma^d_A$.
Moreover, for $\Omega_A=0$, the equality of the first and second moment in \cref{eq:Pm} follows $\tilde{g}\propto g^2_{AB}$ for small parameter values up to slightly above $g^{}_{AB}=\gamma^d_A$, see the solid black curve in \cref{fig:P2_vs_g}(b).
For $\tilde{g}=0$, the equality of the first and second moment implies $\Omega_A\propto \sqrt{g^{}_{AB}}$ for small parameter values up to slightly below $g^{}_{AB}=\gamma^d_A$, see \cref{fig:P2_Om_vs_phiAB}.

\subsection{Frequency synchronization}
Another perspective on these synchronization phenomena is provided by the study of frequency synchronization.
In contrast to before, where we studied the phase synchronization of oscillators, we now compute their oscillation frequencies.
The power spectrum
\begin{align}
    S_{ij\dots}(\omega) &= \lim_{t\to\infty}\int\limits_{-\infty}^{\infty}\d\tau \,C_{ij\dots}(t,\tau)\ee^{\ii\omega\tau}\,,\label{eq:spectrum}
\end{align}
is the Fourier transform of the two-time correlations
\begin{align}
    C_{AA}(t,\tau) &= \langle a^\dagger(t+\tau)a(t)\rangle\,,\\ 
    C_{BB}(t,\tau) &= \langle b^\dagger(t+\tau)b(t)\rangle\,,\\ 
    C_{ABAB}(t,\tau) &= \langle b^\dagger(t+\tau)a(t+\tau)a^\dagger(t)b(t)\rangle\,,
\end{align}
in the steady-state limit $t\to\infty$.
To approximate $S_{AA}$ and $S_{BB}$, we rewrite the Heisenberg equations of motion for $\Omega_A=0$ of the $\tau$-dependent operators as 
\begin{align}
    \frac{\dd}{\dd\tau}\vec{v} = M \vec{v}\,,
\end{align}
where 
\begin{align}
    M &\approx \frac{1}{4}\begin{pmatrix}\gamma^g_A - 2\tilde{g} - 4\gamma^d_A n_A & 2(\ii g^{}_{AB}\ee^{-\ii\phi}-\tilde{g}) \\ 2(\ii g^{}_{AB}\ee^{\ii\phi}-\tilde{g}) & \gamma^g_B-2\tilde{g} - 4\gamma^d_B n_B \end{pmatrix}\,,\\
    \vec{v} &= (\langle a^\dagger(t+\tau)a(t)\rangle, \langle b^\dagger(t+\tau)a(t)\rangle)\,,
\end{align}
and $n_j = \langle a^\dagger_j(t+\tau)a_j(t+\tau)\rangle$.
Here, we approximate $\langle a^{\dagger 2}(t+\tau) a(t+\tau) a(t) \rangle \approx 2\langle a^\dagger(t+\tau) a(t+\tau)\rangle\langle a^\dagger(t+\tau) a(t)\rangle$ using a cumulant expansion of second order and the fact that in the limit $t\to\infty$, i.e., evaluating the expectation values in the steady state, $\langle a^{(\dagger) n}(t+\tau) \rangle=\langle a^{(\dagger) n}(t) \rangle=0$.
For equal rates $\gamma^g_j = \gamma^d_j = \gamma$, the two eigenvalues $\lambda_\pm$ of $M$ read
\begin{align}
    \lambda_\pm =& \frac14(\gamma(1 - 2n_A - 2n_B) - 2\tilde{g})\nonumber\\
    &\pm \frac12\sqrt{\tilde{g}^2 - g^2_{AB} + (n_A - n_B)^2\gamma^2}\,.\label{eq:eigvalsM}
\end{align}
For $n_A \approx n_B$, we can approximate the imaginary part of $\lambda_\pm$ by
\begin{align}
    \omega_\pm = \mathrm{Im}[\lambda_\pm]\approx\pm\sqrt{g^2_{AB} - \tilde{g}^2}/2\,.\label{eq:specapprox}
\end{align}
The correlations $C_{AA}(t,\tau)$ and $C_{BB}(t,\tau)$ effectively measure the time evolution of the phases of the individual oscillators $A$ and $B$.
The correlation $C_{ABAB}(t,\tau)$ is used to obtain the time evolution of the relative phase between both oscillators.
Fourier transforms of these three correlations can be used to distinguish between static and active steady states.
\begin{figure}[t]
    \includegraphics[width=8.6cm]{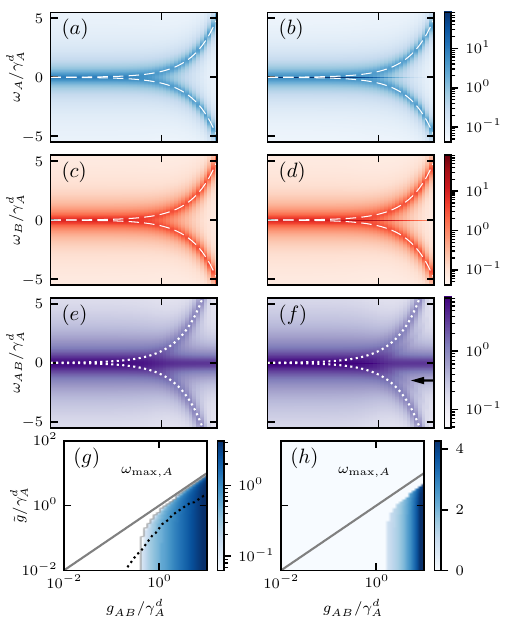}    
    \caption{Power spectra defined in \cref{eq:spectrum} for $\tilde{g}=0.01\gamma^d_A$
    and $\Omega_A=0$ (left column) and $\Omega_A=0.5\gamma^d_A$ (right column).
    (a), (b) $S_{AA}(\omega_A)$. (c), (d) $S_{BB}(\omega_B)$.
    (e), (f) $S_{ABAB}(\omega_{AB})$. (g), (h) Location of maxima of $S_{AA}$. 
    The dashed curves correspond to $\omega_\pm$ of \cref{eq:specapprox}.
    In panels (e) and (f), the white dotted curves correspond to $2\omega_\pm$ and the arrow in (f) points at a local maximum that is close to $\omega_\pm$.
    The black dotted curve in (g) equals the one in \cref{fig:P2_vs_g}(b) and indicates the transition between a single maximum and two maxima in $P_2$.}
    \label{fig:spectrum}
\end{figure}
In \crefs{fig:spectrum}(a) to \ref{fig:spectrum}(f), we present $S_{AA}(\omega_A)$, $S_{BB}(\omega_B)$, and $S_{ABAB}(\omega_{AB})$ for fixed $\tilde{g}=0.01\gamma^d_A$ as a function of $g^{}_{AB}$.
The dashed curves denote the approximation $\omega_\pm$ and the dotted curves in \crefs{fig:spectrum}(e) and \ref{fig:spectrum}(f) denote $2\omega_\pm$.
For $\Omega_A=0.5\gamma^d_A$ (right column), the individual spectra $S_{AA}$ and $S_{BB}$ exhibit an additional local maximum at $\omega_j=0$ ($j=A,B$) that fades out for $g^{}_{AB} \gg \tilde{g}$.
This means that the oscillators have the possibility to lock to the frequency of the drive.
In \cref{fig:spectrum}(f) a local maximum at $\omega_\pm$ is visible (black arrow) which can be interpreted as follows: one of the oscillators locks to the drive while the other one is oscillating at frequency $\omega_\pm$.
We show the location of the maxima of $S_{AA}$ in \crefs{fig:spectrum}(g) and \ref{fig:spectrum}(h).
The dotted curve in \cref{fig:spectrum}(g) is identical to the one in \cref{fig:P2_vs_g}(b) and indicates the transition between a single maximum and two maxima in $P_2$.
Below this curve, the relative phase between the oscillators locks to $\phi_{AB}\approx\pm\pi/2$.
This region of bistable phase locking partially overlaps with the region of frequency locking to nonvanishing $\omega_j$ while the spectrum of the relative frequency has a dominating maximum at $\omega_{AB}=0$. 
This partial overlap may be related to the fact that quantum states lock their phase and frequency only probabilistically: therefore, both effects can occur independently.
In classical systems, states that exhibit a vanishing relative frequency also exhibit locking of their relative phase.
States that feature both frequency locking to $\omega_j\neq0$ and a vanishing relative frequency simultaneously are known as \textit{traveling-wave} states.
We will present exemplary time evolutions of such states in \cref{sec:MF}.
Moreover, in \cref{sec:qtraj}, quantum trajectories of two coherently coupled and undriven oscillators that exhibit antiphase locking and traveling waves are shown.

The relation between phase and frequency locking of traveling-wave states is also analyzed in systems of nonreciprocally coupled groups of multiple spins $1/2$~\cite{PhysRevX.15.011010}.
In \cref{sec:MF}, we present the phase diagram of the mean-field equations of multiple such oscillators.

\begin{figure}[t]
    \centering
    \includegraphics[width=8.6cm]{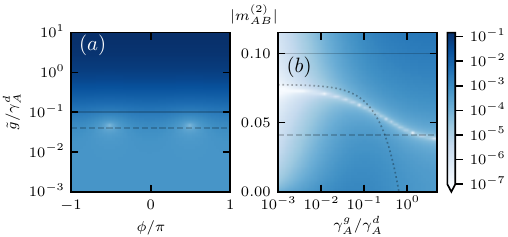}
    \caption{Second moment of the combined synchronization measure $P_2$ for $g^{}_{AB}=0.1\gamma^d_A$ and $\Omega_A=0$. (a) Equal rates $\gamma^g_A=\gamma^g_B=\gamma^d_A=\gamma^d_B$. (b) Different rates $\gamma^g_A=\gamma^g_B$ and $\gamma^d_A=\gamma^d_B$ and $\phi=-\pi/2$.
    The dotted curve corresponds to the approximation defined in \cref{eq:gapproxgamma}.
    In both panels, the solid line denotes $\tilde{g}=g^{}_{AB}$ and the dashed line denotes $\tilde{g}=g^{}_{AB}/\sqrt{6}$.
    The latter expression is obtained from \cref{eq:Pm}.}
    \label{fig:P2_moments_0}
\end{figure}

\section{Blockades}\label{sec:blockades}
If the first-order contribution to the synchronization measure of the relative phase of two coupled oscillators vanishes and the second-order contribution remains, the oscillators are in the so-called synchronization blockade.
Here, since $m^{(1)}_{AB}=0$, bistable locking of their relative phase corresponding to $m^{(2)}_{AB}$ (see the previous sections) is the leading order.
This bistable locking can be interpreted to be mediated by an effective second-order interaction, see \cref{sec:MF_pert}.
Intuitively, information is carried back \textit{and} forth between both oscillators.
Therefore, we would expect the second moment $m^{(2)}_{AB}$ to vanish when at least one of the effective couplings $g^\text{eff}_{A\rightharpoonup B}$ or $g^\text{eff}_{A\leftharpoondown B}$ of \cref{eq:geffAB,eq:geffBA} vanishes: at $\phi=\pm\pi/2$ and $\tilde{g}=g^{}_{AB}$. 
However, this is not the case.
In \cref{fig:P2_moments_0}(a), we show the second moment of the combined synchronization measure $P_2$.
The two zeros of $m^{(2)}_{AB}$ at $\phi=\pm\pi/2$ can be approximated by the dashed gray line that denotes $\tilde{g}=g^{}_{AB}/\sqrt{6}$.
This approximation is based on \cref{eq:Pm}, where the powers were obtained by a perturbation expansion up to third order in $g^{}_{AB}$, $\tilde{g}$, and $\Omega_A$.
The prefactors were extracted from a fit of the maximum of $P_m$ to numerical data presented in \cref{fig:phi_max}(e).

In \cref{fig:P2_moments_0}(b), we show the dependence of the zero of $m^{(2)}_{AB}$ on the ratio $\gamma^g_A/\gamma^d_A$.
Small values of this ratio correspond to the quantum limit, i.e., small radii of the quantum limit cycle meaning small amplitudes of the oscillator.
We expand the steady state of identical oscillators with \textit{different} gain and damping rates $\gamma^g_A=\gamma^g_B$ and $\gamma^d_A=\gamma^d_B$ up to second order in $\tilde{g}/\gamma^d_A$ and $g^{}_{AB}/\gamma^d_A$.
This leads to an approximation of the value of $\tilde{g}$ at which the second moment of the combined synchronization measure vanishes:
for $\gamma^g_A\ll\gamma^d_A$,
\begin{align}
    \tilde{g} \approx \sqrt{\frac35}\left(1 - \frac{3 \gamma^g_A}{40 \gamma^d_A} (12+5\sqrt{3})\right)g^{}_{AB}\,. \label{eq:gapproxgamma}
\end{align}
This approximation is shown in \cref{fig:P2_moments_0}(b) as the dotted curve.

\begin{figure}[t]
    \centering
    \includegraphics[width=8.6cm]{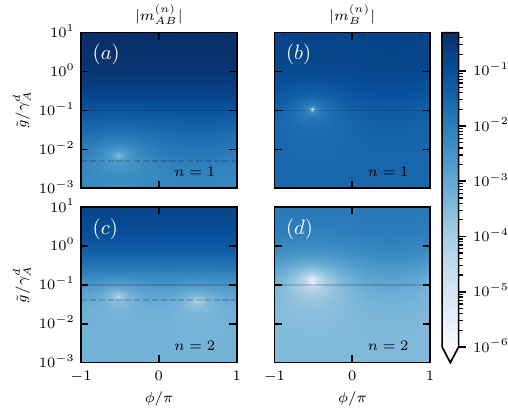}
    \caption{Moments of the combined and individual synchronization measures $P_2(\phi_{AB})$ and $P_1(\phi_{B})$ for $g^{}_{AB}=0.1\gamma^d_A$ and $\Omega_A=0.2\gamma^d_A$. The solid line denotes $\tilde{g}=g^{}_{AB}$. The dashed line in (a) corresponds to the approximation $\tilde{g}=5g^{}_{AB}\Omega_A^2/4$ and the one in (c) to $\tilde{g}=g^{}_{AB}/\sqrt{6}$. Both lines are obtained from \cref{eq:Pm}.}
    \label{fig:P2_moments_1}
\end{figure}

\begin{figure*}[t]
    \centering
    \includegraphics[width=\linewidth]{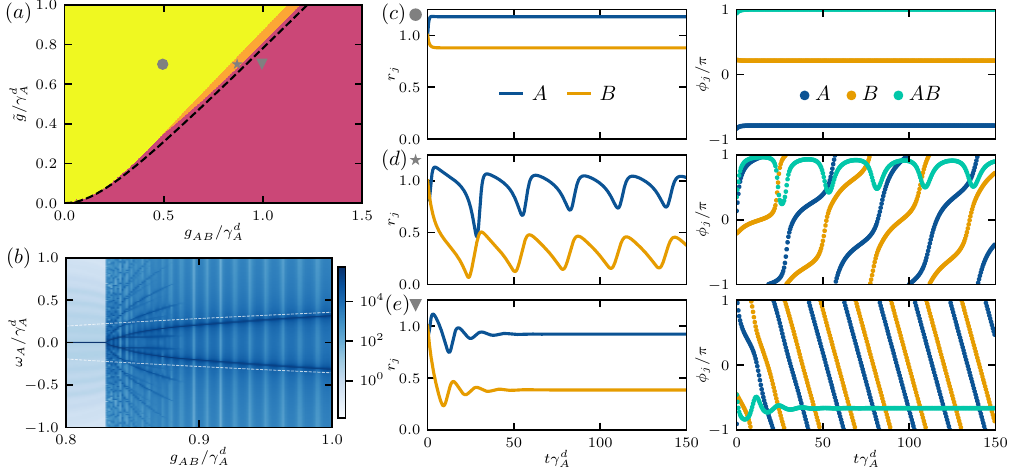}
    \caption{Two oscillators without external drive $\Omega_A=0$ described by \cref{eq:MFa3,eq:MFb3}.
    (a) Phase diagram where each color denotes a different phase.
    The dashed black curve corresponds to the approximate phase boundary between locking and bistable locking, see \cref{eq:pbapprox}.
    (b) Spectrum $S_{AA}(\omega_A)$ for $\tilde{g}=0.7\gamma^d_A$ (location of symbols in (a)).
    The dashed white curve corresponds to $\omega_\pm$ defined in \cref{eq:specapprox}.
    In the regime of modulated traveling-wave states, i.e., the example presented in (d), several maxima exist.
    Panels (c) to (e) show the time evolutions of one phase each corresponding to the symbol next to the panel label.
    The values of $g^{}_{AB}$ and $\tilde{g}$ equal the coordinates of the respective symbol in (a).
    (c) Phase locking to $\phi_{AB}=\pi$.
    (d) Modulated traveling-wave states: varying amplitudes and oscillating phases around $\phi_{AB}\approx \pm\pi/2$.
    (e) Traveling-wave states: constantly increasing phases with fixed $\phi_{AB}\approx \pm\pi/2$.}
    \label{fig:classphasediagram2}
\end{figure*}

More insights into the quantum synchronization mechanisms of unidirectional coupling are obtained by considering an external drive acting on oscillator $A$.
In \cref{fig:P2_moments_1}, we show the first two moments of $P_2(\phi_{AB})$ and $P_1(\phi_B)$.
According to \cref{eq:geffAB}, the influence of the drive on the undriven oscillator $B$ mediated by oscillator $A$ vanishes for $\phi=-\pi/2$ and $g^{}_{AB}=\tilde{g}$, cf.~the zero in \cref{fig:P2_moments_1}(b). 
This can be understood since the Heisenberg equation of motion for $b$ is independent of $a$,
\begin{align}
    \frac{\d}{\d t}b = \frac{\gamma^g_B-2\tilde{g}}{4} b - \frac{\gamma^d_B}{2} b^\dagger b^2\,,\label{eq:HEOMb1}
\end{align}
cf.~\cref{eq:HEOMb}.
Analogously, \cref{eq:HEOMb1} is invariant under the U(1) transformation $b \to \ee^{\ii\theta}b$ such that oscillator $B$ shows no phase preference. 
However, the relative phase between $A$ and $B$ as well as the phase of $A$ is locked. 
These effects can be understood intuitively by imagining a quantum trajectory of these \textit{unidirectionally} interacting oscillators.
Oscillator $B$ evolves independently from $A$, but $A$ is influenced by (the random jumps of) $B$.
Thus, the relative phase $\phi_{AB}$ is locked even if $\phi_B$ is not.
In this way, $B$ can be interpreted as an additional noise source acting on $A$.

At fixed $g^{}_{AB}$, $\phi=-\pi/2$, and $\Omega_A\neq0$, increasing $\tilde{g}$ leads to switches from locking ($\phi_{AB}=0$) to bistable locking and back to locking ($\phi_{AB}=\pi$), cf.~\cref{fig:phi_max}(c).
At some value close to $\tilde{g}=5g^{}_{AB}\Omega_A^2/4$, indicated by the dashed line in \cref{fig:P2_moments_1}(a), the relative phase between both oscillators exhibits bistable locking even if both phases lock to a single value individually.
This approximation is obtained from \cref{eq:Pm}.
In comparison to \cref{fig:P2_moments_0}(b), the minima of the second moment $m^{(2)}_{AB}$ shown in \cref{fig:P2_moments_1}(c) lie at different values of $\tilde{g}$: in the presence of the external drive the symmetry between $\phi=-\pi/2$ and $\phi=\pi/2$ is broken.
A perturbation expansion of the first and second moment of the synchronization measure of the undriven oscillator $B$ to leading order in $g^{}_{AB}$, $\tilde{g}$, and $\Omega_A$ yields
\begin{align}
    m^{(1)}_B &=  u_5\frac{\ii\tilde{g} - \ee^{-\ii\phi} g^{}_{AB}}{\gamma^2}\Omega_A\,,\\
    m^{(2)}_B &= -(\ii u_6\tilde{g} + u_7 \ee^{-\ii\phi} g^{}_{AB})\frac{\ii\tilde{g} - \ee^{-\ii\phi}g^{}_{AB}}{\gamma^4}\Omega_A^2\,.
\end{align}
Both equations suggest a zero at $\tilde{g}=g^{}_{AB}$ and $\phi=-\pi/2$.
Within this approximation, the second zero in \cref{fig:P2_moments_1}(d) can be explained by opposite signs of $u_6$ and $u_7$.

\begin{figure*}[t]
    \centering
    \includegraphics[width=\linewidth]{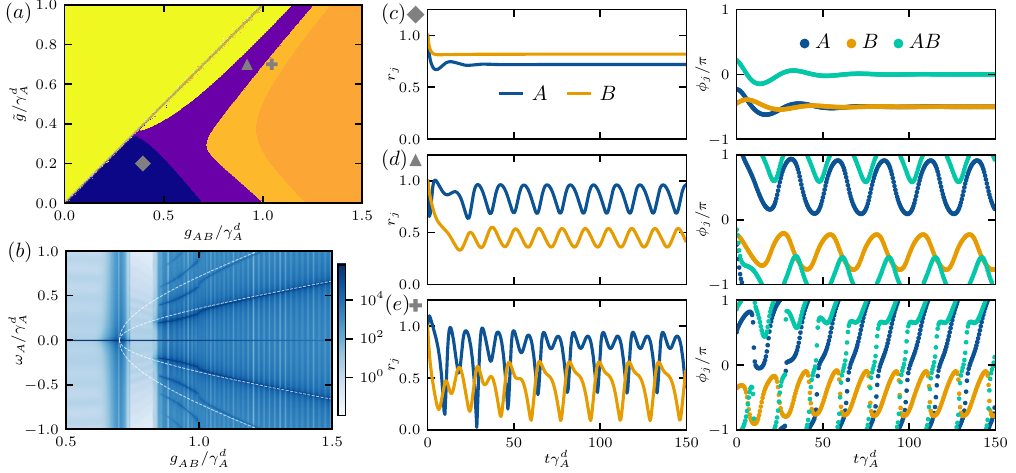}
    \caption{Two oscillators with external drive $\Omega_A = 0.5\gamma^g_A$ described by \cref{eq:MFa3,eq:MFb3}.
    (a) Phase diagram where each color denotes a different phase.
    White pixels were not assigned any phase.
    The gray line denotes $\tilde{g}=g^{}_{AB}$.
    (b) Spectrum $S_{AA}(\omega_A)$ for $\tilde{g}=0.7\gamma^d_A$ (location of upper symbols in (a)).
    The dashed white curves indicate $\omega_\pm$ and $2\omega_\pm$ defined in \cref{eq:specapprox}.
    Panels (c) to (e) show the time evolutions of one phase each corresponding to the symbol next to the panel label.
    The values of $g^{}_{AB}$ and $\tilde{g}$ equal the coordinates of the respective symbol in (a).
    (c) Phase locking to $\phi_A=\phi_B=-\pi/2$.
    (d) Wobble motion: varying amplitudes and oscillating phases around $\phi_{AB}\approx \pm\pi/2$.
    (e) Partial traveling-wave states: constantly increasing $\phi_A$, oscillating $\phi_{B}$ around $-\pi/2$.
    The yellow phase (top left) corresponds to phase locking to $\phi_{AB}=\pi$, similar to \cref{fig:classphasediagram2}(c), where for $\tilde{g}>g^{}_{AB}$ ($\tilde{g}<g^{}_{AB}$) $\phi_A=-\pi/2$ ($\phi_A=\pi/2$).
    The darker orange phase (center right) hosts modulated traveling-wave states, similar to \cref{fig:classphasediagram2}(d).
    Videos of time evolutions are provided in~\cite{SuppMat}.}
    \label{fig:classphasediagram2drive}
\end{figure*}

\begin{figure*}[t]
    \centering
    \includegraphics[width=\linewidth]{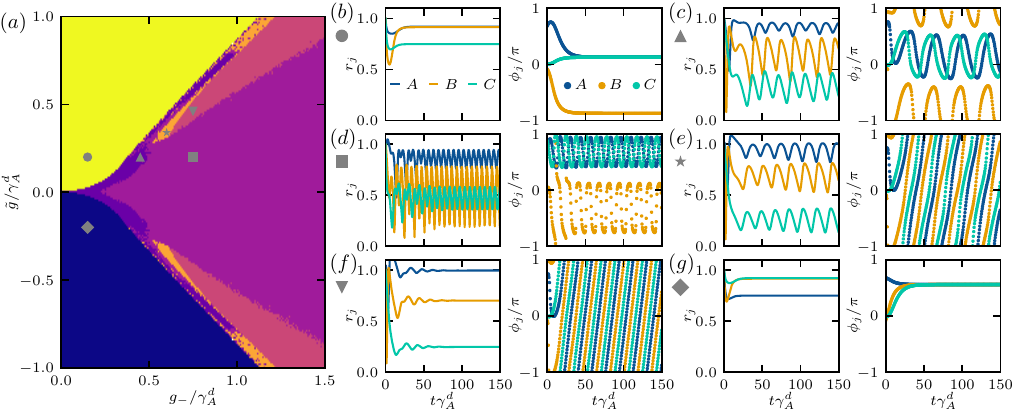}
    \caption{Open chain of three oscillators described by \cref{eq:EOM3osc}.
    (a) Phase diagram where each color denotes a different phase.
    White pixels were not assigned any phase.
    Panels (b) to (g) show the time evolutions of one phase each corresponding to the symbol next to the panel label.
    The values of $g_-$ and $\tilde{g}$ equal the coordinates of the respective symbol in (a).
    (b) Phase locking to $\Delta\phi_j=\phi_{j}-\phi_{j+1}=\pi$.
    (c) Wobble motion: varying amplitudes and oscillating phases around $\Delta\phi_j\approx\pi$.
    (d) Both wobble motion (c) and traveling-wave states (f) exist.
    (e) Modulated traveling-wave states: varying amplitudes and oscillating phases around $\Delta\phi_j\approx \pm 2\pi/3$.
    (f) Traveling-wave states: constantly increasing phases with fixed $\Delta\phi_j \approx \pm 2\pi/3$.
    (g) Phase locking to $\Delta\phi_j=0$.
    Videos of time evolutions are provided in~\cite{SuppMat}.}
    \label{fig:classphasediagram3}
\end{figure*}

\section{Classical Analogue}\label{sec:MF}
In this section, as a comparison to the phase diagram of the relative phase between the two quantum oscillators shown in \cref{fig:phi_max}(b), we will discuss the phase diagrams of the classical analogues of two and three quantum oscillators.

\subsection{Two oscillators}
The phase diagrams of the relative phase between two quantum oscillators in the mean-field limit are obtained from the equations 
\begin{align}
    \langle \dot{a}\rangle &= -\ii\frac{\Omega_A}{2} -\frac{\ii g^{}_{AB}\ee^{\ii\phi} + \tilde{g}}{2} \langle b\rangle +\frac{\gamma^g_A-2\tilde{g} - 2\gamma^d_A|\langle a\rangle|^2}{4}\langle a\rangle\label{eq:MFa3}\,,\\
    \langle \dot{b}\rangle &= -\frac{\ii g^{}_{AB}\ee^{-\ii\phi} + \tilde{g}}{2} \langle a\rangle+\frac{\gamma^g_B-2\tilde{g} - 2\gamma^d_B|\langle b\rangle|^2}{4}\langle b\rangle \label{eq:MFb3}\,,
\end{align}
see \cref{sec:effectivemodel2}.
These equations have been studied in the context of exceptional points~\cite{weis2023}.
The phase diagram for $\Omega_A=0$ is presented in \cref{fig:classphasediagram2}(a). 
As in the previous sections, we consider identical oscillators $\gamma^g_A=\gamma^g_B$ and $\gamma^d_A=\gamma^d_B$ as well as $\phi=-\pi/2$.
To avoid vanishing linear gain that would lead to both oscillators collapsing to zero amplitude, we fix $\gamma^g_A - 2\tilde{g}= \gamma^d_A$.
We identify the following regimes: (i) phase locking to $\phi_{AB}=\pi$, (ii) phase locking to $\phi_{AB}=0$, (iii) traveling-wave states with $\phi_{AB}\approx \pm\pi/2$, and (iv) modulated traveling-wave states.
If $g^{}_{AB}=0$ and $\tilde{g}>0$ both oscillators want to lock to the other oscillator with $\phi_{AB}=\pi$.
For small $g^{}_{AB}\neq0$ we expect bistable locking for $\tilde{g}$ smaller than $2g^2_{AB}/\gamma^g_A$, see \cref{sec:MF_pert}, resulting in the boundary
\begin{align}
    \tilde{g} = \sqrt{g^2_{AB} + (\gamma^d_A/4)^2} - \frac{\gamma^d_A}{4}\,.\label{eq:pbapprox}
\end{align}
This boundary corresponds to the dashed black curve in \cref{fig:classphasediagram2}(a).
The especially interesting so-called (modulated) traveling-wave states are identified by bistable locking of their relative phase and  monotonic growing phases of oscillation.
Traveling waves exhibit fixed amplitudes and modulated traveling waves exhibit varying amplitudes.
Such active states have been studied in the context of nonreciprocal phase transitions~\cite{Fruchart2021,PhysRevX.14.011029,PhysRevX.15.011010}.
The spectra of oscillators in such states show maxima at nonvanishing frequencies, see \cref{fig:classphasediagram2}(b).
Note that in the regime of modulated traveling waves maxima at frequencies lower than the expected oscillation frequency appear.
These are very likely related to the modulation frequencies of the variation of the amplitude and oscillation frequency.

The phase diagram of two oscillators for $\Omega_A=0.5\gamma^d_A$ is shown in \cref{fig:classphasediagram2drive}(a).
Here, in addition to the regions of (i) locking to $\phi_{AB}=\pi$ (yellow, top left) and (ii) modulated traveling-wave states (darker orange, center right) known from \cref{fig:classphasediagram2}, we find: (iii) locking to $\phi_A=\phi_B=-\pi/2$, (iv) wobble motion, and (v) partial traveling-wave states.
The wobble motion is identified by varying phases $\phi_j$ within an interval smaller than $2\pi$ as well as varying amplitude, see \cref{fig:classphasediagram2drive}(d).
Our definition of the wobble motion also includes states that are assigned to the so-called swap phase discussed in~\cite{Fruchart2021}.
In the swap phase, the oscillators are aligned on a line and periodically switch between a static $\phi_j$ and $\phi_j+\pi$.
To distinguish between the wobble motion and traveling waves, we use the following order parameter
\begin{align}
    S_{\text{ori},j} = \frac{1}{\tau}\int\limits_{T}^{T+\tau}\d t\,\text{sign}(\text{Im}[\langle \dot{a}_j \rangle \langle a_j \rangle^* ])\,.
\end{align}
The integrand measures the orientation of rotation which is averaged over a time interval $\tau$ when the steady state is reached ($T\gamma^d_A \gg 1$).
If the state switches between clockwise and counterclockwise rotation, i.e., varying phase around fixed values (wobble motion), $|S_{\text{ori},j}|$ will be small.
However, if a state does not change its orientation of rotation, $|S_{\text{ori},j}|$ will be close to unity.
This is the case for (modulated) traveling waves.
In \cref{fig:classphasediagram2drive}(e), we present an example trajectory of partial traveling-wave states.
Here, only oscillator $A$ performs full rotations ($|S_{\text{ori},A}|\approx 1$), whereas oscillator $B$ is still in a wobble motion ($S_{\text{ori},A}\approx 0$).
Remarkably, in this phase and for this choice of nonreciprocal coupling ($\phi=-\pi/2$), the undriven oscillator $B$ is more localized to the phase $\phi_B \approx -\pi/2$ induced by the drive than the driven oscillator which is rotating monotonically.
A similar behavior is seen in the spectra of the quantum oscillators in \crefs{fig:spectrum}(b) and \ref{fig:spectrum}(d), where the peak at $\omega_A=0$ is less dominant than the peak at $\omega_B=0$.
In the Supplemental Material~\cite{SuppMat}, we provide videos of time evolutions for each phase that are shown in \crefs{fig:classphasediagram2drive}(c--e).

\subsection{Three oscillators}
We also consider the next more complex system consisting of three oscillators.
In \cref{fig:classphasediagram3}(a), we present the phase diagram of an open chain of oscillators that obey
\begin{align}
    \langle \dot{a}_j\rangle =& \frac{\gamma^g_j-2\tilde{g} - 2\gamma^d_j|\langle a_j\rangle|^2}{4}\langle a_j\rangle \nonumber\\
    &-\frac{G^{}_{j,j+1}}{2} \langle a_{j+1}\rangle -\frac{G^{}_{j,j-1}}{2} \langle a_{j-1}\rangle\,.\label{eq:EOM3osc}
\end{align}
Here, we set $\gamma^d_j=\gamma^d_A$ and fix $\gamma^g_j-2\tilde{g}=\gamma^d_A$.
The couplings
\begin{align}
    G^{}_{j,j+1} &= G^{}_{j,(j+1)\,\text{mod}\,3} = \tilde{g} + g_-\,,\\
    G^{}_{j,j-1} &= G^{}_{j,(j-1)\,\text{mod}\,3} = \tilde{g} - g_-\,,
\end{align}
are chosen identical for each oscillator and $g_-$ corresponds to $\ii g^{}_{AB}\ee^{\ii\phi}$ and $\phi=-\pi/2$.

In the open chain, $G^{}_{C,A}=G^{}_{A,C}=0$ vanish.
The phase diagram of the open chain is rich: (i) phase locking to $\Delta\phi_j=\phi_{j}-\phi_{j+1}=\pi$, (ii) phase locking to $\Delta\phi_j=0$, (iii) traveling waves, (iv) modulated traveling waves, (v) wobble motion, and (vi) both wobble motion and traveling waves.
To distinguish states performing the wobble motion and fully rotating traveling-wave states, in addition to $S_{\text{ori},j}$, we employ the following order parameter
\begin{align}
    S_{\text{rot},j} =  \left| \frac{1}{\tau}\int\limits_{T}^{T+\tau}\d t\,\ee^{\ii\phi_j} \right| \,.
\end{align}
It is the magnitude of a time average of the complex phase factors when the steady state is reached ($T\gamma^d_A \gg 1$). 
The order parameter $S_{\text{rot},j}$ reaches values close to zero for fully rotating (modulated) traveling-wave states, values close to unity for static states, and values in between for states performing a wobble motion.
For each pixel in \cref{fig:classphasediagram3}(a), we generate time evolutions of $100$ random initial states $\langle a_j\rangle = \exp(\ii\phi_j)$ that are drawn from a uniform distribution over the interval $\phi_j\in[0,2\pi]$.
In the Supplemental Material~\cite{SuppMat}, we provide videos of time evolutions for each phase that are shown in \crefs{fig:classphasediagram3}(b--g).

\section{Conclusion}
We have investigated the interplay of three phase-locking mechanisms of two quantum limit-cycle oscillators induced by an external drive, a coherent coupling, and a dissipative coupling leading to three different steady states.
In this setup, the effective nonreciprocal interaction can be tuned to be unidirectional.
For vanishing drive strength and increasing nonreciprocity, the following sequence of events occurs: (i) interaction terms in the mean-field equations become unidirectional, (ii) the second moment of the combined quantum synchronization measure vanishes, (iii) a switch from phase locking to bistable locking occurs, and (iv) the first and second moment of the combined quantum synchronization measure become equal. 
Varying the drive strength of an external signal acting on one of the two oscillators, the magnitude of a coherent coupling, and the strength of a dissipative interaction, we have shown that the steady-state value of the relative phase between the oscillators can be tuned.
Making use of the quantum synchronization measure evaluated for a perturbation expansion of the steady state in the three parameters drive strength, coherent coupling, and dissipative interaction, we have qualitatively explained the transitions between the three regimes of phase localization.
This perturbation expansion has been used to identify minima of the magnitude of the second moment of the synchronization measure of the relative phase. 
Moreover, regions of bistable locking partially overlap with regions in which two-time correlations exhibit a periodic time dependence similar to traveling-wave states.
Such traveling-wave states have also been found as steady-state solutions of the mean-field approximation of the master equation of the quantum system.
For two and three noreciprocally coupled oscillators in the mean-field limit, we have found highly nontrivial active states by defining suitable order parameters. 

Nonreciprocity in (open) quantum systems and their classical analogues is a rapidly emerging field in nonlinear quantum physics.
Future research directions include the study of (frustrated) networks of $N \geq 3$ quantum oscillators as well as their (potentially existing) nonreciprocal phase transitions.

\section*{Acknowledgements}
We would like to thank Clara Wanjura, Tobias Nadolny, and Parvinder Solanki for fruitful discussions.
We acknowledge financial support from the Swiss National Science Foundation individual grant (Grant No. 200020 200481).
We furthermore acknowledge the use of \textsc{QuTip}~\cite{QuTiP} and \textsc{multiprocess}~\cite{multiprocess}.
The data that support the findings of this article are openly available~\cite{data}.

\appendix

\section{Quantum Synchronization Measure}\label{sec:synch_measures}
To obtain a better intuition for the operator $\tilde{a}$ introduced in \cref{eq:atilde} that resembles the Susskind-Glogower operator~\cite{PhysicsPhysiqueFizika.1.49}, we compute its expectation value in a coherent state $\ket{\alpha}$,
\begin{align}
    \bra{\alpha}\tilde{a}^k\ket{\alpha} &= \alpha^k \,\ee^{-|\alpha|^2}\sum_{n=0}^\infty\frac{|\alpha|^{2n}}{n!\sqrt{(n+1)\dots(n+k)}}\,.
\end{align}
The two limiting behaviors are
\begin{align}
    \bra{\alpha}\tilde{a}^k\ket{\alpha}\stackrel{|\alpha|\ll 1}{\approx}& \frac{\alpha^k}{\sqrt{k!}}\,,\\
    \bra{\alpha}\tilde{a}^k\ket{\alpha}\stackrel{|\alpha|\gg 1}{\approx}& \left(\frac{\alpha}{|\alpha|}\right)^k\,.\label{eq:atildelarge}
\end{align}
If $|\alpha|\gg 1$, $\tilde{a}$ effectively becomes a phase operator that measures the phase $\phi_0$ of a coherent state $\ket{\alpha=r\ee^{\ii\phi_0}}$. 
Therefore, for coherent states, the absolute value of $\langle \tilde{a}\rangle$ is upper bounded by the absolute value of the quantum synchronization measure introduced in~\cite{Weiss_2016},
\begin{align}
    S=\frac{\langle a\rangle}{\sqrt{\langle a^\dagger a\rangle}}\,.
\end{align}
For a coherent state $\ket{\alpha=0}$ that is considered to show no form of quantum synchronization, $\langle\tilde{a}^k\rangle$ vanishes whereas $S\neq 0$. 
The synchronization measure $P_1$ defined in \cref{eq:P1viaOp} can be approximated as
\begin{align}
    P_1(\phi) \stackrel{|\alpha|\ll 1}{\approx}& \frac{|\alpha|}{\pi}\cos(\phi_0-\phi)\,,\\
    P_1(\phi) \stackrel{|\alpha|\gg 1}{\approx}& \frac{1}{2\pi}\sum_{k=-\infty}^\infty\ee^{\ii k(\phi_0-\phi)} - \frac{1}{2\pi}=\delta(\phi_0-\phi)-\frac{1}{2\pi}\,.
\end{align}
Thus, for $|\alpha|\ll 1$, the first moment of $P_1$ dominates, whereas for $|\alpha|\gg 1$ the phase distribution $P_1$ can be expressed by a Dirac $\delta$ distribution.

\section{Effective Model}\label{sec:effectivemodel}
In this appendix, we focus on the effective two-oscillator quantum model studied in the main text.
The dissipative coupling between two quantum van der Pol oscillators discussed in \cref{sec:dissipative} can be realized by introducing an auxiliary rapidly decaying cavity~\cite{PhysRevX.5.021025}.

\subsection{Three Quantum Oscillators}\label{sec:threeosc}
The Lindblad master equation of the full three-oscillator model reads
\begin{align}
    \dot{\rho} =&  -\ii[H, \rho] + \tilde{\mathcal{L}}(\rho)\,,
\end{align}
with
\begin{align}
    H =& \frac{\Omega_A}{2}a^\dagger + \frac{g^{}_{AB}}{2}\ee^{\ii\phi} a^\dagger b + \frac{g}{2} (b^\dagger c + c^\dagger a) + \mathrm{H.c.}\,,
\end{align}
and
\begin{align}
    \tilde{\mathcal{L}}(\rho) =& \frac{\kappa}{2}\mathcal{D}[c](\rho) + \frac{\gamma^g_A}{2}\mathcal{D}[a^\dagger](\rho) + \frac{\gamma^g_B}{2}\mathcal{D}[b^\dagger](\rho)\nonumber\\
    &+ \frac{\gamma^d_A}{2}\mathcal{D}[a^2](\rho) + \frac{\gamma^d_B}{2}\mathcal{D}[b^2](\rho)\,.\label{eq:MEQ3osc}
\end{align}
The two quantum van der Pol oscillators that are denoted by the annihilation operators $a$ and $b$ are coherently coupled with strength $g^{}_{AB}$ and phase $\phi$.
The gain and damping rates are defined as $\gamma^g_j$ and $\gamma^d_j$.
Oscillator $A$ is driven by an external drive $\Omega_A$.
Furthermore, both oscillators are coherently coupled with strength $g$ to a linearly decaying cavity characterized by operators $c$ whose decay rate $\kappa$ is significantly larger than any other timescale of the system.

The Heisenberg equation of motion of the cavity operator $c$ reads
\begin{align}
    \frac{\d}{\d t} c = -\ii \frac{g}{2}(a+b)-\frac{\kappa}{4}c\,.
\end{align}
For $\kappa\gg\gamma^d_j,\gamma^g_j$ we can assume that the cavity reaches its steady state much faster than oscillator $A$ and $B$.
Therefore, we replace $c \to -2\ii (a+b) g/\kappa$ in \cref{eq:MEQ3osc} leading to
\begin{align}
    \frac{g}{2}(b^\dagger c + c^\dagger a) + \mathrm{H.c.} \to 0\,,\\
    \frac{\kappa}{2}\mathcal{D}[c](\rho) \to 2\frac{g^2}{\kappa}\mathcal{D}[a+b](\rho) \,.
\end{align}
In this limit, the system can be described effectively by two quantum van der Pol oscillators interacting dissipatively.

\subsection{Two Quantum Oscillators}\label{sec:effectivemodel2}
As described in the previous subsection, the effective Hamiltonian and dissipators for a rapidly decaying cavity read
\begin{align}
    H_\text{eff} =& \frac{\Omega_A}{2}a^\dagger + \frac{g^{}_{AB}}{2}\ee^{\ii\phi} a^\dagger b + \mathrm{H.c.}\\
    \tilde{\mathcal{L}}(\rho) =& \tilde{g}\mathcal{D}[a+b](\rho) + \frac{\gamma^g_A}{2}\mathcal{D}[a^\dagger](\rho) + \frac{\gamma^g_B}{2}\mathcal{D}[b^\dagger](\rho)\nonumber\\
    &+ \frac{\gamma^d_A}{2}\mathcal{D}[a^2](\rho) + \frac{\gamma^d_B}{2}\mathcal{D}[b^2](\rho)\,,
\end{align}
where $\tilde{g}=2g^2/\kappa$.
The resulting Heisenberg equations of motion are
\begin{align}
    \frac{\d}{\d t}a &= -\ii\frac{\Omega_A}{2}\mathds{1} -\frac{\ii g^{}_{AB}\ee^{\ii\phi} + \tilde{g}}{2} b +\frac{\gamma^g_A-2\tilde{g}}{4} a - \frac{\gamma^d_A}{2} a^\dagger a^2 \label{eq:HEOMa}\,,\\
    \frac{\d}{\d t} b &= -\frac{\ii g^{}_{AB}\ee^{-\ii\phi} + \tilde{g}}{2} a + \frac{\gamma^g_B-2\tilde{g}}{4} b - \frac{\gamma^d_B}{2} b^\dagger b^2 \label{eq:HEOMb}\,.
\end{align}
A cumulant expansion to lowest order yields the mean-field equations \cref{eq:MFa3,eq:MFb3}.
In this parametrization, the coupling between $A$ and $B$ becomes unidirectional for $g^{}_{AB}=\tilde{g}$ and $\phi=\pm\pi/2$.
By inserting $\langle a_j\rangle=r_j\ee^{\ii\phi_j}$, we obtain
\begin{align}
    \dot{r}_A =& -\frac{\Omega_A}{2}\sin(\phi_A)+\frac{\gamma^g_A-2\tilde{g}}{4}r_A-\frac{\gamma^d_A}{2}r_A^3\nonumber\\
        &-\frac{r_B}{2}(g^{}_{AB}\sin(\phi_{AB}-\phi)+\tilde{g}\cos(\phi_{AB}))\,, \\
    \dot{r}_B =& \frac{\gamma^g_B-2\tilde{g}}{4}r_B-\frac{\gamma^d_B}{2}r_B^3\nonumber\\
        &+\frac{r_A}{2}(g^{}_{AB}\sin(\phi_{AB}-\phi)-\tilde{g}\cos(\phi_{AB}))\,,
\end{align}
as well as
\begin{align}
    \dot{\phi}_A =& -\frac{\Omega_A}{2r_A}\cos(\phi_A)\nonumber\\
        &- \frac{r_B}{2r_A}(g^{}_{AB}\cos(\phi_{AB}-\phi)-\tilde{g}\sin(\phi_{AB}))\,, \\
    \dot{\phi}_B =& - \frac{r_A}{2r_B}(g^{}_{AB}\cos(\phi_{AB}-\phi)+\tilde{g}\sin(\phi_{AB})) \\
    \dot{\phi}_{AB} =& -\frac{\Omega_A}{2r_A}\cos(\phi_A) +\frac{\tilde{g}}{2}\left(\frac{r_A}{r_B}+\frac{r_B}{r_A}\right)\sin(\phi_{AB})\nonumber\\
        &+\frac{g^{}_{AB}}{2}\left(\frac{r_A}{r_B}-\frac{r_B}{r_A}\right)\cos(\phi_{AB}-\phi)\,.    
\end{align}

\subsection{Approximate Steady-State Solution}\label{sec:MF_pert}
The perturbative solution of the steady-state radii
\begin{align}
    r_j=r^{(0)}_j+\epsilon r^{(1)}_j=\sqrt{\frac{\gamma^g_j}{2\gamma^d_j}}+\epsilon r^{(1)}_j\,,
\end{align}
for $\Omega_A=0$ reads 
\begin{align}
    \epsilon r^{(1)}_A =& -\frac{\tilde{g}}{\gamma^g_A}(r^{(0)}_A+r^{(0)}_B\cos(\phi_{AB})) \nonumber\\
    &-\frac{g^{}_{AB}}{\gamma^g_A}r^{(0)}_B\sin(\phi_{AB}-\phi)\,,\\
    \epsilon r^{(1)}_B =& -\frac{\tilde{g}}{\gamma^g_B}(r^{(0)}_B+r^{(0)}_A\cos(\phi_{AB})) \nonumber\\
    &+\frac{g^{}_{AB}}{\gamma^g_B}r^{(0)}_A\sin(\phi_{AB}-\phi)\,.
\end{align}
For identical oscillators $\gamma^g_A=\gamma^g_B$ and $\gamma^d_A=\gamma^d_B$, this leads to
\begin{align}
    \dot{\phi}_{AB} =& \tilde{g}\sin(\phi_{AB}) - \frac{g^2_{AB}}{\gamma^g_A}\sin(2(\phi_{AB}-\phi))\,.
\end{align}
If $\tilde{g}\gg g^2_{AB}/\gamma^g_A$, a single stable solution $\phi_{AB}=\pi$ exists. 
If $\tilde{g}\ll g^2_{AB}/\gamma^g_A$, the system experiences bistable locking to $\phi_{AB} = \phi,\phi+\pi $.
For $\phi=\pm\pi/2$, there are two stable solutions $\phi_{AB}=\pm\arccos(-\tilde{g}\gamma^g_A/2g^2_{AB})$ if $\tilde{g}<2g^2_{AB}/\gamma^g_A$ and there is a single stable solution $\phi_{AB}=\pi$ if $\tilde{g}>2g^2_{AB}/\gamma^g_A$.

\begin{figure}[b]
    \centering
    \includegraphics[width=8.6cm]{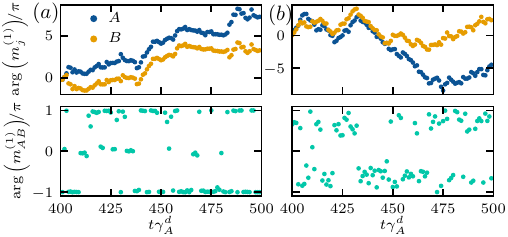}
    \caption{Quantum trajectories of two coherently coupled oscillators corresponding to \cref{fig:P2_vs_g}.
    We set $\Omega_A=0$ and $\tilde{g}=0.01\gamma^d_A$.
    (a) Antiphase locking to $\phi_{AB}\approx\pi$ for $g^{}_{AB}=0.1\gamma^d_A$.
    (b) Bistable phase locking to $\phi_{AB}\approx\pm\pi/2$ for $g^{}_{AB}=2\gamma^d_A$.
    In the upper row, the complex phases are unwrapped: the difference between subsequent values never exceeds $\pm\pi$ due to added shifts of $\pm2\pi$.}
    \label{fig:trajectories_no_drive}
\end{figure}

\section{Quantum Trajectories}\label{sec:qtraj}
In addition to the steady-state analysis of the density matrix presented in \cref{fig:P2_vs_g}, we simulate individual quantum trajectories.
Following~\cite{Wiseman2009}, the stochastic quantum master equation reads
\begin{align}
    \d\rho_m =& -\ii[H,\rho_m]\d t + \tilde{\mathcal{L}}(\rho_m)\d t + \tilde{g}\mathcal{D}[a+b](\rho_m)\d t\nonumber\\
    &+ \sqrt{\tilde{g}}[(a+b - \Tr{(a+b)\rho_m})\rho_m + \mathrm{H.c.}] \d W\,,
\end{align}
where the first line describes the deterministic part with $H=g^{}_{AB}\ee^{\ii\phi}a^\dagger b/2  + \mathrm{H.c.}$ and $\tilde{\mathcal{L}}$ is defined in \cref{eq:Leom}.
In the second line, the stochastic part with Wiener increment $\d W$ originates from the dissipative interaction $\tilde{g}\mathcal{D}[a+b]$.
As described in \cref{sec:threeosc}, this interaction is mediated by a lossy cavity.
Monitoring the signal leaking out of this cavity leads to insights about the expectation value $\langle a+b \rangle$ that caries information about the relative phase $\phi_{AB}$.
The density matrix $\rho_m$ is the state conditioned on the outcome of a measurement of $\langle a+b \rangle$.
Numerically, we compute various operator expectation values using $\rho_m$.
For the the case $\Omega_A=0$, we compute the first moments $m^{(1)}_j$ of the individual operators as well as the first moment of the combined synchronization measure $m^{(2)}_{AB}$.
Their complex argument effectively corresponds to $\phi_j$ and $\phi_{AB}$ and is shown in \cref{fig:trajectories_no_drive} for $\tilde{g}=0.01\gamma^d_A$.
For $g^{}_{AB}=0.1\gamma^d_A$ the relative phase locks to $\phi_{AB}\approx\pi$, whereas for $g^{}_{AB}=2\gamma^d_A$, bistable locking to $\phi_{AB}\approx\pm\pi/2$ occurs.
In \cref{fig:trajectories_no_drive}(b), one can furthermore identify the traveling-wave character, i.e., linearly increasing/decreasing phases $\phi_j$, as well as a correlation between the signs of $\dot{\phi}_j$ and $\phi_{AB}$.
The quantum trajectories presented in \cref{fig:trajectories_no_drive} should be compared with the steady-state analysis shown in \crefs{fig:P2_vs_g}(b) and \ref{fig:spectrum}(g).


\begin{thebibliography}{44}%
	\makeatletter
	\providecommand \@ifxundefined [1]{%
		\@ifx{#1\undefined}
	}%
	\providecommand \@ifnum [1]{%
		\ifnum #1\expandafter \@firstoftwo
		\else \expandafter \@secondoftwo
		\fi
	}%
	\providecommand \@ifx [1]{%
		\ifx #1\expandafter \@firstoftwo
		\else \expandafter \@secondoftwo
		\fi
	}%
	\providecommand \natexlab [1]{#1}%
	\providecommand \enquote  [1]{``#1''}%
	\providecommand \bibnamefont  [1]{#1}%
	\providecommand \bibfnamefont [1]{#1}%
	\providecommand \citenamefont [1]{#1}%
	\providecommand \href@noop [0]{\@secondoftwo}%
	\providecommand \href [0]{\begingroup \@sanitize@url \@href}%
	\providecommand \@href[1]{\@@startlink{#1}\@@href}%
	\providecommand \@@href[1]{\endgroup#1\@@endlink}%
	\providecommand \@sanitize@url [0]{\catcode `\\12\catcode `\$12\catcode
		`\&12\catcode `\#12\catcode `\^12\catcode `\_12\catcode `\%12\relax}%
	\providecommand \@@startlink[1]{}%
	\providecommand \@@endlink[0]{}%
	\providecommand \url  [0]{\begingroup\@sanitize@url \@url }%
	\providecommand \@url [1]{\endgroup\@href {#1}{\urlprefix }}%
	\providecommand \urlprefix  [0]{URL }%
	\providecommand \Eprint [0]{\href }%
	\providecommand \doibase [0]{https://doi.org/}%
	\providecommand \selectlanguage [0]{\@gobble}%
	\providecommand \bibinfo  [0]{\@secondoftwo}%
	\providecommand \bibfield  [0]{\@secondoftwo}%
	\providecommand \translation [1]{[#1]}%
	\providecommand \BibitemOpen [0]{}%
	\providecommand \bibitemStop [0]{}%
	\providecommand \bibitemNoStop [0]{.\EOS\space}%
	\providecommand \EOS [0]{\spacefactor3000\relax}%
	\providecommand \BibitemShut  [1]{\csname bibitem#1\endcsname}%
	\let\auto@bib@innerbib\@empty
	%</preamble>
	\bibitem [{\citenamefont {Pikovsky}\ \emph {et~al.}(2001)\citenamefont
		{Pikovsky}, \citenamefont {Rosenblum},\ and\ \citenamefont
		{Kurths}}]{Synch_Pikovsky}%
	\BibitemOpen
	\bibfield  {author} {\bibinfo {author} {\bibfnamefont {A.}~\bibnamefont
			{Pikovsky}}, \bibinfo {author} {\bibfnamefont {M.}~\bibnamefont
			{Rosenblum}},\ and\ \bibinfo {author} {\bibfnamefont {J.}~\bibnamefont
			{Kurths}},\ }\href {https://doi.org/10.1017/cbo9780511755743} {\emph
		{\bibinfo {title} {{Synchronization: A Universal Concept in Nonlinear
					Science}}}}\ (\bibinfo  {publisher} {Cambridge University Press, Cambridge,
		England},\ \bibinfo {year} {2001})\BibitemShut {NoStop}%
	\bibitem [{\citenamefont {Strogatz}(2003)}]{Synch_Strogatz}%
	\BibitemOpen
	\bibfield  {author} {\bibinfo {author} {\bibfnamefont {S.~H.}\ \bibnamefont
			{Strogatz}},\ }\href@noop {} {\emph {\bibinfo {title} {{Sync: The Emerging
					Science of Spontaneous Order}}}}\ (\bibinfo  {publisher} {Hyperion, New
		York},\ \bibinfo {year} {2003})\BibitemShut {NoStop}%
	\bibitem [{\citenamefont {Acebr\'on}\ \emph {et~al.}(2005)\citenamefont
		{Acebr\'on}, \citenamefont {Bonilla}, \citenamefont {P\'erez~Vicente},
		\citenamefont {Ritort},\ and\ \citenamefont {Spigler}}]{RevModPhys.77.137}%
	\BibitemOpen
	\bibfield  {author} {\bibinfo {author} {\bibfnamefont {J.~A.}\ \bibnamefont
			{Acebr\'on}}, \bibinfo {author} {\bibfnamefont {L.~L.}\ \bibnamefont
			{Bonilla}}, \bibinfo {author} {\bibfnamefont {C.~J.}\ \bibnamefont
			{P\'erez~Vicente}}, \bibinfo {author} {\bibfnamefont {F.}~\bibnamefont
			{Ritort}},\ and\ \bibinfo {author} {\bibfnamefont {R.}~\bibnamefont
			{Spigler}},\ }\bibfield  {title} {\bibinfo {title} {The {Kuramoto} model: A
			simple paradigm for synchronization phenomena},\ }\href
	{https://doi.org/10.1103/RevModPhys.77.137} {\bibfield  {journal} {\bibinfo
			{journal} {Rev. Mod. Phys.}\ }\textbf {\bibinfo {volume} {77}},\ \bibinfo
		{pages} {137} (\bibinfo {year} {2005})}\BibitemShut {NoStop}%
	\bibitem [{\citenamefont {Balanov}\ \emph {et~al.}(2008)\citenamefont
		{Balanov}, \citenamefont {Janson}, \citenamefont {Postnov},\ and\
		\citenamefont {Sosnovtseva}}]{Balanov2008}%
	\BibitemOpen
	\bibfield  {author} {\bibinfo {author} {\bibfnamefont {A.}~\bibnamefont
			{Balanov}}, \bibinfo {author} {\bibfnamefont {N.}~\bibnamefont {Janson}},
		\bibinfo {author} {\bibfnamefont {D.}~\bibnamefont {Postnov}},\ and\ \bibinfo
		{author} {\bibfnamefont {O.}~\bibnamefont {Sosnovtseva}},\ }\href
	{https://doi.org/10.1007/978-3-540-72128-4} {\emph {\bibinfo {title}
			{{Synchronization - From Simple to Complex}}}}\ (\bibinfo  {publisher}
	{Springer Science \& Business Media},\ \bibinfo {address} {Berlin
		Heidelberg},\ \bibinfo {year} {2008})\BibitemShut {NoStop}%
	\bibitem [{\citenamefont {Ludwig}\ and\ \citenamefont
		{Marquardt}(2013)}]{PhysRevLett.111.073603}%
	\BibitemOpen
	\bibfield  {author} {\bibinfo {author} {\bibfnamefont {M.}~\bibnamefont
			{Ludwig}}\ and\ \bibinfo {author} {\bibfnamefont {F.}~\bibnamefont
			{Marquardt}},\ }\bibfield  {title} {\bibinfo {title} {{Quantum Many-Body
				Dynamics in Optomechanical Arrays}},\ }\href
	{https://doi.org/10.1103/PhysRevLett.111.073603} {\bibfield  {journal}
		{\bibinfo  {journal} {Phys. Rev. Lett.}\ }\textbf {\bibinfo {volume} {111}},\
		\bibinfo {pages} {073603} (\bibinfo {year} {2013})}\BibitemShut {NoStop}%
	\bibitem [{\citenamefont {Lee}\ and\ \citenamefont
		{Sadeghpour}(2013)}]{Synch_vdP_Lee}%
	\BibitemOpen
	\bibfield  {author} {\bibinfo {author} {\bibfnamefont {T.~E.}\ \bibnamefont
			{Lee}}\ and\ \bibinfo {author} {\bibfnamefont {H.~R.}\ \bibnamefont
			{Sadeghpour}},\ }\bibfield  {title} {\bibinfo {title} {Quantum
			synchronization of quantum van der {Pol} oscillators with trapped ions},\
	}\href {https://doi.org/10.1103/PhysRevLett.111.234101} {\bibfield  {journal}
		{\bibinfo  {journal} {Phys. Rev. Lett.}\ }\textbf {\bibinfo {volume} {111}},\
		\bibinfo {pages} {234101} (\bibinfo {year} {2013})}\BibitemShut {NoStop}%
	\bibitem [{\citenamefont {Walter}\ \emph {et~al.}(2015)\citenamefont {Walter},
		\citenamefont {Nunnenkamp},\ and\ \citenamefont {Bruder}}]{Synch_vdP_Walter}%
	\BibitemOpen
	\bibfield  {author} {\bibinfo {author} {\bibfnamefont {S.}~\bibnamefont
			{Walter}}, \bibinfo {author} {\bibfnamefont {A.}~\bibnamefont {Nunnenkamp}},\
		and\ \bibinfo {author} {\bibfnamefont {C.}~\bibnamefont {Bruder}},\
	}\bibfield  {title} {\bibinfo {title} {Quantum synchronization of two van der
			{Pol} oscillators},\ }\href
	{https://doi.org/https://doi.org/10.1002/andp.201400144} {\bibfield
		{journal} {\bibinfo  {journal} {Annalen der Physik}\ }\textbf {\bibinfo
			{volume} {527}},\ \bibinfo {pages} {131} (\bibinfo {year}
		{2015})}\BibitemShut {NoStop}%
	\bibitem [{\citenamefont {Chia}\ \emph {et~al.}(2020)\citenamefont {Chia},
		\citenamefont {Kwek},\ and\ \citenamefont {Noh}}]{Chia2020}%
	\BibitemOpen
	\bibfield  {author} {\bibinfo {author} {\bibfnamefont {A.}~\bibnamefont
			{Chia}}, \bibinfo {author} {\bibfnamefont {L.~C.}\ \bibnamefont {Kwek}},\
		and\ \bibinfo {author} {\bibfnamefont {C.}~\bibnamefont {Noh}},\ }\bibfield
	{title} {\bibinfo {title} {Relaxation oscillations and frequency entrainment
			in quantum mechanics},\ }\href {https://doi.org/10.1103/PhysRevE.102.042213}
	{\bibfield  {journal} {\bibinfo  {journal} {Phys. Rev. E}\ }\textbf {\bibinfo
			{volume} {102}},\ \bibinfo {pages} {042213} (\bibinfo {year}
		{2020})}\BibitemShut {NoStop}%
	\bibitem [{\citenamefont {Ben~Arosh}\ \emph {et~al.}(2021)\citenamefont
		{Ben~Arosh}, \citenamefont {Cross},\ and\ \citenamefont
		{Lifshitz}}]{PhysRevResearch.3.013130}%
	\BibitemOpen
	\bibfield  {author} {\bibinfo {author} {\bibfnamefont {L.}~\bibnamefont
			{Ben~Arosh}}, \bibinfo {author} {\bibfnamefont {M.~C.}\ \bibnamefont
			{Cross}},\ and\ \bibinfo {author} {\bibfnamefont {R.}~\bibnamefont
			{Lifshitz}},\ }\bibfield  {title} {\bibinfo {title} {{Quantum limit cycles
				and the Rayleigh and van der Pol oscillators}},\ }\href
	{https://doi.org/10.1103/PhysRevResearch.3.013130} {\bibfield  {journal}
		{\bibinfo  {journal} {Phys. Rev. Res.}\ }\textbf {\bibinfo {volume} {3}},\
		\bibinfo {pages} {013130} (\bibinfo {year} {2021})}\BibitemShut {NoStop}%
	\bibitem [{\citenamefont {Roulet}\ and\ \citenamefont
		{Bruder}(2018{\natexlab{a}})}]{PhysRevLett.121.053601}%
	\BibitemOpen
	\bibfield  {author} {\bibinfo {author} {\bibfnamefont {A.}~\bibnamefont
			{Roulet}}\ and\ \bibinfo {author} {\bibfnamefont {C.}~\bibnamefont
			{Bruder}},\ }\bibfield  {title} {\bibinfo {title} {{Synchronizing the
				Smallest Possible System}},\ }\href
	{https://doi.org/10.1103/PhysRevLett.121.053601} {\bibfield  {journal}
		{\bibinfo  {journal} {Phys. Rev. Lett.}\ }\textbf {\bibinfo {volume} {121}},\
		\bibinfo {pages} {053601} (\bibinfo {year} {2018}{\natexlab{a}})}\BibitemShut
	{NoStop}%
	\bibitem [{\citenamefont {Roulet}\ and\ \citenamefont
		{Bruder}(2018{\natexlab{b}})}]{PhysRevLett.121.063601}%
	\BibitemOpen
	\bibfield  {author} {\bibinfo {author} {\bibfnamefont {A.}~\bibnamefont
			{Roulet}}\ and\ \bibinfo {author} {\bibfnamefont {C.}~\bibnamefont
			{Bruder}},\ }\bibfield  {title} {\bibinfo {title} {{Quantum Synchronization
				and Entanglement Generation}},\ }\href
	{https://doi.org/10.1103/PhysRevLett.121.063601} {\bibfield  {journal}
		{\bibinfo  {journal} {Phys. Rev. Lett.}\ }\textbf {\bibinfo {volume} {121}},\
		\bibinfo {pages} {063601} (\bibinfo {year} {2018}{\natexlab{b}})}\BibitemShut
	{NoStop}%
	\bibitem [{\citenamefont {Parra-L\'opez}\ and\ \citenamefont
		{Bergli}(2020)}]{PhysRevA.101.062104}%
	\BibitemOpen
	\bibfield  {author} {\bibinfo {author} {\bibfnamefont {A.}~\bibnamefont
			{Parra-L\'opez}}\ and\ \bibinfo {author} {\bibfnamefont {J.}~\bibnamefont
			{Bergli}},\ }\bibfield  {title} {\bibinfo {title} {Synchronization in
			two-level quantum systems},\ }\href
	{https://doi.org/10.1103/PhysRevA.101.062104} {\bibfield  {journal} {\bibinfo
			{journal} {Phys. Rev. A}\ }\textbf {\bibinfo {volume} {101}},\ \bibinfo
		{pages} {062104} (\bibinfo {year} {2020})}\BibitemShut {NoStop}%
	\bibitem [{\citenamefont {Ameri}\ \emph {et~al.}(2015)\citenamefont {Ameri},
		\citenamefont {Eghbali-Arani}, \citenamefont {Mari}, \citenamefont {Farace},
		\citenamefont {Kheirandish}, \citenamefont {Giovannetti},\ and\ \citenamefont
		{Fazio}}]{PhysRevA.91.012301}%
	\BibitemOpen
	\bibfield  {author} {\bibinfo {author} {\bibfnamefont {V.}~\bibnamefont
			{Ameri}}, \bibinfo {author} {\bibfnamefont {M.}~\bibnamefont
			{Eghbali-Arani}}, \bibinfo {author} {\bibfnamefont {A.}~\bibnamefont {Mari}},
		\bibinfo {author} {\bibfnamefont {A.}~\bibnamefont {Farace}}, \bibinfo
		{author} {\bibfnamefont {F.}~\bibnamefont {Kheirandish}}, \bibinfo {author}
		{\bibfnamefont {V.}~\bibnamefont {Giovannetti}},\ and\ \bibinfo {author}
		{\bibfnamefont {R.}~\bibnamefont {Fazio}},\ }\bibfield  {title} {\bibinfo
		{title} {Mutual information as an order parameter for quantum
			synchronization},\ }\href {https://doi.org/10.1103/PhysRevA.91.012301}
	{\bibfield  {journal} {\bibinfo  {journal} {Phys. Rev. A}\ }\textbf {\bibinfo
			{volume} {91}},\ \bibinfo {pages} {012301} (\bibinfo {year}
		{2015})}\BibitemShut {NoStop}%
	\bibitem [{\citenamefont {Mari}\ \emph {et~al.}(2013)\citenamefont {Mari},
		\citenamefont {Farace}, \citenamefont {Didier}, \citenamefont {Giovannetti},\
		and\ \citenamefont {Fazio}}]{PhysRevLett.111.103605}%
	\BibitemOpen
	\bibfield  {author} {\bibinfo {author} {\bibfnamefont {A.}~\bibnamefont
			{Mari}}, \bibinfo {author} {\bibfnamefont {A.}~\bibnamefont {Farace}},
		\bibinfo {author} {\bibfnamefont {N.}~\bibnamefont {Didier}}, \bibinfo
		{author} {\bibfnamefont {V.}~\bibnamefont {Giovannetti}},\ and\ \bibinfo
		{author} {\bibfnamefont {R.}~\bibnamefont {Fazio}},\ }\bibfield  {title}
	{\bibinfo {title} {{Measures of Quantum Synchronization in Continuous
				Variable Systems}},\ }\href {https://doi.org/10.1103/PhysRevLett.111.103605}
	{\bibfield  {journal} {\bibinfo  {journal} {Phys. Rev. Lett.}\ }\textbf
		{\bibinfo {volume} {111}},\ \bibinfo {pages} {103605} (\bibinfo {year}
		{2013})}\BibitemShut {NoStop}%
	\bibitem [{\citenamefont {Manzano}\ \emph {et~al.}(2013)\citenamefont
		{Manzano}, \citenamefont {Galve}, \citenamefont {Giorgi}, \citenamefont
		{Hern{\'a}ndez-Garc{\'\i}a},\ and\ \citenamefont
		{Zambrini}}]{manzano2013synchronization}%
	\BibitemOpen
	\bibfield  {author} {\bibinfo {author} {\bibfnamefont {G.}~\bibnamefont
			{Manzano}}, \bibinfo {author} {\bibfnamefont {F.}~\bibnamefont {Galve}},
		\bibinfo {author} {\bibfnamefont {G.~L.}\ \bibnamefont {Giorgi}}, \bibinfo
		{author} {\bibfnamefont {E.}~\bibnamefont {Hern{\'a}ndez-Garc{\'\i}a}},\ and\
		\bibinfo {author} {\bibfnamefont {R.}~\bibnamefont {Zambrini}},\ }\bibfield
	{title} {\bibinfo {title} {Synchronization, quantum correlations and
			entanglement in oscillator networks},\ }\href
	{https://doi.org/10.1038/srep01439} {\bibfield  {journal} {\bibinfo
			{journal} {Scientific Reports}\ }\textbf {\bibinfo {volume} {3}},\ \bibinfo
		{pages} {1439} (\bibinfo {year} {2013})}\BibitemShut {NoStop}%
	\bibitem [{\citenamefont {L\"orch}\ \emph {et~al.}(2017)\citenamefont
		{L\"orch}, \citenamefont {Nigg}, \citenamefont {Nunnenkamp}, \citenamefont
		{Tiwari},\ and\ \citenamefont {Bruder}}]{PhysRevLett.118.243602}%
	\BibitemOpen
	\bibfield  {author} {\bibinfo {author} {\bibfnamefont {N.}~\bibnamefont
			{L\"orch}}, \bibinfo {author} {\bibfnamefont {S.~E.}\ \bibnamefont {Nigg}},
		\bibinfo {author} {\bibfnamefont {A.}~\bibnamefont {Nunnenkamp}}, \bibinfo
		{author} {\bibfnamefont {R.~P.}\ \bibnamefont {Tiwari}},\ and\ \bibinfo
		{author} {\bibfnamefont {C.}~\bibnamefont {Bruder}},\ }\bibfield  {title}
	{\bibinfo {title} {{Quantum Synchronization Blockade: Energy Quantization
				Hinders Synchronization of Identical Oscillators}},\ }\href
	{https://doi.org/10.1103/PhysRevLett.118.243602} {\bibfield  {journal}
		{\bibinfo  {journal} {Phys. Rev. Lett.}\ }\textbf {\bibinfo {volume} {118}},\
		\bibinfo {pages} {243602} (\bibinfo {year} {2017})}\BibitemShut {NoStop}%
	\bibitem [{\citenamefont {Solanki}\ \emph {et~al.}(2023)\citenamefont
		{Solanki}, \citenamefont {Mehdi}, \citenamefont {Hajdu\ifmmode~\check{s}\else
			\v{s}\fi{}ek},\ and\ \citenamefont {Vinjanampathy}}]{PhysRevA.108.022216}%
	\BibitemOpen
	\bibfield  {author} {\bibinfo {author} {\bibfnamefont {P.}~\bibnamefont
			{Solanki}}, \bibinfo {author} {\bibfnamefont {F.~M.}\ \bibnamefont {Mehdi}},
		\bibinfo {author} {\bibfnamefont {M.}~\bibnamefont
			{Hajdu\ifmmode~\check{s}\else \v{s}\fi{}ek}},\ and\ \bibinfo {author}
		{\bibfnamefont {S.}~\bibnamefont {Vinjanampathy}},\ }\bibfield  {title}
	{\bibinfo {title} {Symmetries and synchronization blockade},\ }\href
	{https://doi.org/10.1103/PhysRevA.108.022216} {\bibfield  {journal} {\bibinfo
			{journal} {Phys. Rev. A}\ }\textbf {\bibinfo {volume} {108}},\ \bibinfo
		{pages} {022216} (\bibinfo {year} {2023})}\BibitemShut {NoStop}%
	\bibitem [{\citenamefont {Kehrer}\ \emph {et~al.}(2024)\citenamefont {Kehrer},
		\citenamefont {Nadolny},\ and\ \citenamefont {Bruder}}]{PhysRevA.110.042203}%
	\BibitemOpen
	\bibfield  {author} {\bibinfo {author} {\bibfnamefont {T.}~\bibnamefont
			{Kehrer}}, \bibinfo {author} {\bibfnamefont {T.}~\bibnamefont {Nadolny}},\
		and\ \bibinfo {author} {\bibfnamefont {C.}~\bibnamefont {Bruder}},\
	}\bibfield  {title} {\bibinfo {title} {Quantum synchronization through the
			interference blockade},\ }\href {https://doi.org/10.1103/PhysRevA.110.042203}
	{\bibfield  {journal} {\bibinfo  {journal} {Phys. Rev. A}\ }\textbf {\bibinfo
			{volume} {110}},\ \bibinfo {pages} {042203} (\bibinfo {year}
		{2024})}\BibitemShut {NoStop}%
	\bibitem [{\citenamefont {Ivlev}\ \emph {et~al.}(2015)\citenamefont {Ivlev},
		\citenamefont {Bartnick}, \citenamefont {Heinen}, \citenamefont {Du},
		\citenamefont {Nosenko},\ and\ \citenamefont {L\"owen}}]{PhysRevX.5.011035}%
	\BibitemOpen
	\bibfield  {author} {\bibinfo {author} {\bibfnamefont {A.~V.}\ \bibnamefont
			{Ivlev}}, \bibinfo {author} {\bibfnamefont {J.}~\bibnamefont {Bartnick}},
		\bibinfo {author} {\bibfnamefont {M.}~\bibnamefont {Heinen}}, \bibinfo
		{author} {\bibfnamefont {C.-R.}\ \bibnamefont {Du}}, \bibinfo {author}
		{\bibfnamefont {V.}~\bibnamefont {Nosenko}},\ and\ \bibinfo {author}
		{\bibfnamefont {H.}~\bibnamefont {L\"owen}},\ }\bibfield  {title} {\bibinfo
		{title} {{Statistical Mechanics where Newton's Third Law is Broken}},\ }\href
	{https://doi.org/10.1103/PhysRevX.5.011035} {\bibfield  {journal} {\bibinfo
			{journal} {Phys. Rev. X}\ }\textbf {\bibinfo {volume} {5}},\ \bibinfo {pages}
		{011035} (\bibinfo {year} {2015})}\BibitemShut {NoStop}%
	\bibitem [{\citenamefont {Ramaswamy}(2010)}]{Ramaswamy2010}%
	\BibitemOpen
	\bibfield  {author} {\bibinfo {author} {\bibfnamefont {S.}~\bibnamefont
			{Ramaswamy}},\ }\bibfield  {title} {\bibinfo {title} {{The Mechanics and
				Statistics of Active Matter}},\ }\href
	{https://doi.org/10.1146/annurev-conmatphys-070909-104101} {\bibfield
		{journal} {\bibinfo  {journal} {Annual Review of Condensed Matter Physics}\
		}\textbf {\bibinfo {volume} {1}},\ \bibinfo {pages} {323} (\bibinfo {year}
		{2010})}\BibitemShut {NoStop}%
	\bibitem [{\citenamefont {Schweitzer}(2019)}]{Schweitzer2019}%
	\BibitemOpen
	\bibfield  {author} {\bibinfo {author} {\bibfnamefont {F.}~\bibnamefont
			{Schweitzer}},\ }\bibfield  {title} {\bibinfo {title} {{An agent-based
				framework of active matter with applications in biological and social
				systems}},\ }\href {https://doi.org/10.1088/1361-6404/aaeb63} {\bibfield
		{journal} {\bibinfo  {journal} {European Journal of Physics}\ }\textbf
		{\bibinfo {volume} {40}},\ \bibinfo {pages} {014003} (\bibinfo {year}
		{2019})}\BibitemShut {NoStop}%
	\bibitem [{\citenamefont {Lotka}(1920)}]{Lotka1920}%
	\BibitemOpen
	\bibfield  {author} {\bibinfo {author} {\bibfnamefont {A.~J.}\ \bibnamefont
			{Lotka}},\ }\bibfield  {title} {\bibinfo {title} {{Analytical Note on Certain
				Rhythmic Relations in Organic Systems}},\ }\href
	{https://doi.org/10.1073/pnas.6.7.410} {\bibfield  {journal} {\bibinfo
			{journal} {Proceedings of the National Academy of Sciences}\ }\textbf
		{\bibinfo {volume} {6}},\ \bibinfo {pages} {410–415} (\bibinfo {year}
		{1920})}\BibitemShut {NoStop}%
	\bibitem [{\citenamefont {Volterra}(1926)}]{VOLTERRA1926}%
	\BibitemOpen
	\bibfield  {author} {\bibinfo {author} {\bibfnamefont {V.}~\bibnamefont
			{Volterra}},\ }\bibfield  {title} {\bibinfo {title} {{Fluctuations in the
				Abundance of a Species considered Mathematically\textsuperscript{1}}},\
	}\href {https://doi.org/10.1038/118558a0} {\bibfield  {journal} {\bibinfo
			{journal} {Nature}\ }\textbf {\bibinfo {volume} {118}},\ \bibinfo {pages}
		{558–560} (\bibinfo {year} {1926})}\BibitemShut {NoStop}%
	\bibitem [{\citenamefont {Bacaër}(2011)}]{Bacaer2011}%
	\BibitemOpen
	\bibfield  {author} {\bibinfo {author} {\bibfnamefont {N.}~\bibnamefont
			{Bacaër}},\ }\href {https://doi.org/10.1007/978-0-85729-115-8} {\emph
		{\bibinfo {title} {{A Short History of Mathematical Population Dynamics}}}}\
	(\bibinfo  {publisher} {Springer London},\ \bibinfo {year}
	{2011})\BibitemShut {NoStop}%
	\bibitem [{\citenamefont {Fruchart}\ \emph {et~al.}(2021)\citenamefont
		{Fruchart}, \citenamefont {Hanai}, \citenamefont {Littlewood},\ and\
		\citenamefont {Vitelli}}]{Fruchart2021}%
	\BibitemOpen
	\bibfield  {author} {\bibinfo {author} {\bibfnamefont {M.}~\bibnamefont
			{Fruchart}}, \bibinfo {author} {\bibfnamefont {R.}~\bibnamefont {Hanai}},
		\bibinfo {author} {\bibfnamefont {P.~B.}\ \bibnamefont {Littlewood}},\ and\
		\bibinfo {author} {\bibfnamefont {V.}~\bibnamefont {Vitelli}},\ }\bibfield
	{title} {\bibinfo {title} {Non-reciprocal phase transitions},\ }\href
	{https://doi.org/10.1038/s41586-021-03375-9} {\bibfield  {journal} {\bibinfo
			{journal} {Nature}\ }\textbf {\bibinfo {volume} {592}},\ \bibinfo {pages}
		{363} (\bibinfo {year} {2021})}\BibitemShut {NoStop}%
	\bibitem [{\citenamefont {Hanai}(2024)}]{PhysRevX.14.011029}%
	\BibitemOpen
	\bibfield  {author} {\bibinfo {author} {\bibfnamefont {R.}~\bibnamefont
			{Hanai}},\ }\bibfield  {title} {\bibinfo {title} {{Nonreciprocal Frustration:
				Time Crystalline Order-by-Disorder Phenomenon and a Spin-Glass-like State}},\
	}\href {https://doi.org/10.1103/PhysRevX.14.011029} {\bibfield  {journal}
		{\bibinfo  {journal} {Phys. Rev. X}\ }\textbf {\bibinfo {volume} {14}},\
		\bibinfo {pages} {011029} (\bibinfo {year} {2024})}\BibitemShut {NoStop}%
	\bibitem [{\citenamefont {Hatano}\ and\ \citenamefont
		{Nelson}(1996)}]{PhysRevLett.77.570}%
	\BibitemOpen
	\bibfield  {author} {\bibinfo {author} {\bibfnamefont {N.}~\bibnamefont
			{Hatano}}\ and\ \bibinfo {author} {\bibfnamefont {D.~R.}\ \bibnamefont
			{Nelson}},\ }\bibfield  {title} {\bibinfo {title} {Localization transitions
			in non-hermitian quantum mechanics},\ }\href
	{https://doi.org/10.1103/PhysRevLett.77.570} {\bibfield  {journal} {\bibinfo
			{journal} {Phys. Rev. Lett.}\ }\textbf {\bibinfo {volume} {77}},\ \bibinfo
		{pages} {570} (\bibinfo {year} {1996})}\BibitemShut {NoStop}%
	\bibitem [{\citenamefont {Roth}\ and\ \citenamefont
		{Hammerer}(2016)}]{PhysRevA.94.043841}%
	\BibitemOpen
	\bibfield  {author} {\bibinfo {author} {\bibfnamefont {A.}~\bibnamefont
			{Roth}}\ and\ \bibinfo {author} {\bibfnamefont {K.}~\bibnamefont
			{Hammerer}},\ }\bibfield  {title} {\bibinfo {title} {Synchronization of
			active atomic clocks via quantum and classical channels},\ }\href
	{https://doi.org/10.1103/PhysRevA.94.043841} {\bibfield  {journal} {\bibinfo
			{journal} {Phys. Rev. A}\ }\textbf {\bibinfo {volume} {94}},\ \bibinfo
		{pages} {043841} (\bibinfo {year} {2016})}\BibitemShut {NoStop}%
	\bibitem [{\citenamefont {Lorenzo}\ \emph {et~al.}(2022)\citenamefont
		{Lorenzo}, \citenamefont {Militello}, \citenamefont {Napoli}, \citenamefont
		{Zambrini},\ and\ \citenamefont {Palma}}]{Lorenzo2022}%
	\BibitemOpen
	\bibfield  {author} {\bibinfo {author} {\bibfnamefont {S.}~\bibnamefont
			{Lorenzo}}, \bibinfo {author} {\bibfnamefont {B.}~\bibnamefont {Militello}},
		\bibinfo {author} {\bibfnamefont {A.}~\bibnamefont {Napoli}}, \bibinfo
		{author} {\bibfnamefont {R.}~\bibnamefont {Zambrini}},\ and\ \bibinfo
		{author} {\bibfnamefont {G.~M.}\ \bibnamefont {Palma}},\ }\bibfield  {title}
	{\bibinfo {title} {Quantum synchronisation and clustering in chiral
			networks},\ }\href {https://doi.org/10.1088/1367-2630/ac51a9} {\bibfield
		{journal} {\bibinfo  {journal} {New Journal of Physics}\ }\textbf {\bibinfo
			{volume} {24}},\ \bibinfo {pages} {023030} (\bibinfo {year}
		{2022})}\BibitemShut {NoStop}%
	\bibitem [{\citenamefont {Wanjura}\ \emph {et~al.}(2020)\citenamefont
		{Wanjura}, \citenamefont {Brunelli},\ and\ \citenamefont
		{Nunnenkamp}}]{Wanjura2020}%
	\BibitemOpen
	\bibfield  {author} {\bibinfo {author} {\bibfnamefont {C.~C.}\ \bibnamefont
			{Wanjura}}, \bibinfo {author} {\bibfnamefont {M.}~\bibnamefont {Brunelli}},\
		and\ \bibinfo {author} {\bibfnamefont {A.}~\bibnamefont {Nunnenkamp}},\
	}\bibfield  {title} {\bibinfo {title} {Topological framework for directional
			amplification in driven-dissipative cavity arrays},\ }\href
	{https://doi.org/10.1038/s41467-020-16863-9} {\bibfield  {journal} {\bibinfo
			{journal} {Nature Communications}\ }\textbf {\bibinfo {volume} {11}},\
		\bibinfo {pages} {3149} (\bibinfo {year} {2020})}\BibitemShut {NoStop}%
	\bibitem [{\citenamefont {W\"achtler}\ and\ \citenamefont
		{Platero}(2023)}]{PhysRevResearch.5.023021}%
	\BibitemOpen
	\bibfield  {author} {\bibinfo {author} {\bibfnamefont {C.~W.}\ \bibnamefont
			{W\"achtler}}\ and\ \bibinfo {author} {\bibfnamefont {G.}~\bibnamefont
			{Platero}},\ }\bibfield  {title} {\bibinfo {title} {{Topological
				synchronization of quantum van der Pol oscillators}},\ }\href
	{https://doi.org/10.1103/PhysRevResearch.5.023021} {\bibfield  {journal}
		{\bibinfo  {journal} {Phys. Rev. Res.}\ }\textbf {\bibinfo {volume} {5}},\
		\bibinfo {pages} {023021} (\bibinfo {year} {2023})}\BibitemShut {NoStop}%
	\bibitem [{\citenamefont {Nadolny}\ \emph {et~al.}(2025)\citenamefont
		{Nadolny}, \citenamefont {Bruder},\ and\ \citenamefont
		{Brunelli}}]{PhysRevX.15.011010}%
	\BibitemOpen
	\bibfield  {author} {\bibinfo {author} {\bibfnamefont {T.}~\bibnamefont
			{Nadolny}}, \bibinfo {author} {\bibfnamefont {C.}~\bibnamefont {Bruder}},\
		and\ \bibinfo {author} {\bibfnamefont {M.}~\bibnamefont {Brunelli}},\
	}\bibfield  {title} {\bibinfo {title} {Nonreciprocal synchronization of
			active quantum spins},\ }\href {https://doi.org/10.1103/PhysRevX.15.011010}
	{\bibfield  {journal} {\bibinfo  {journal} {Phys. Rev. X}\ }\textbf {\bibinfo
			{volume} {15}},\ \bibinfo {pages} {011010} (\bibinfo {year}
		{2025})}\BibitemShut {NoStop}%
	\bibitem [{\citenamefont {Metelmann}\ and\ \citenamefont
		{Clerk}(2015)}]{PhysRevX.5.021025}%
	\BibitemOpen
	\bibfield  {author} {\bibinfo {author} {\bibfnamefont {A.}~\bibnamefont
			{Metelmann}}\ and\ \bibinfo {author} {\bibfnamefont {A.~A.}\ \bibnamefont
			{Clerk}},\ }\bibfield  {title} {\bibinfo {title} {Nonreciprocal photon
			transmission and amplification via reservoir engineering},\ }\href
	{https://doi.org/10.1103/PhysRevX.5.021025} {\bibfield  {journal} {\bibinfo
			{journal} {Phys. Rev. X}\ }\textbf {\bibinfo {volume} {5}},\ \bibinfo {pages}
		{021025} (\bibinfo {year} {2015})}\BibitemShut {NoStop}%
	\bibitem [{\citenamefont {Hush}\ \emph {et~al.}(2015)\citenamefont {Hush},
		\citenamefont {Li}, \citenamefont {Genway}, \citenamefont {Lesanovsky},\ and\
		\citenamefont {Armour}}]{phase_dist_Hush}%
	\BibitemOpen
	\bibfield  {author} {\bibinfo {author} {\bibfnamefont {M.~R.}\ \bibnamefont
			{Hush}}, \bibinfo {author} {\bibfnamefont {W.}~\bibnamefont {Li}}, \bibinfo
		{author} {\bibfnamefont {S.}~\bibnamefont {Genway}}, \bibinfo {author}
		{\bibfnamefont {I.}~\bibnamefont {Lesanovsky}},\ and\ \bibinfo {author}
		{\bibfnamefont {A.~D.}\ \bibnamefont {Armour}},\ }\bibfield  {title}
	{\bibinfo {title} {Spin correlations as a probe of quantum synchronization in
			trapped-ion phonon lasers},\ }\href
	{https://doi.org/10.1103/PhysRevA.91.061401} {\bibfield  {journal} {\bibinfo
			{journal} {Phys. Rev. A}\ }\textbf {\bibinfo {volume} {91}},\ \bibinfo
		{pages} {061401(R)} (\bibinfo {year} {2015})}\BibitemShut {NoStop}%
	\bibitem [{\citenamefont {Weiss}\ \emph {et~al.}(2016)\citenamefont {Weiss},
		\citenamefont {Kronwald},\ and\ \citenamefont {Marquardt}}]{Weiss_2016}%
	\BibitemOpen
	\bibfield  {author} {\bibinfo {author} {\bibfnamefont {T.}~\bibnamefont
			{Weiss}}, \bibinfo {author} {\bibfnamefont {A.}~\bibnamefont {Kronwald}},\
		and\ \bibinfo {author} {\bibfnamefont {F.}~\bibnamefont {Marquardt}},\
	}\bibfield  {title} {\bibinfo {title} {Noise-induced transitions in
			optomechanical synchronization},\ }\href
	{https://doi.org/10.1088/1367-2630/18/1/013043} {\bibfield  {journal}
		{\bibinfo  {journal} {New Journal of Physics}\ }\textbf {\bibinfo {volume}
			{18}},\ \bibinfo {pages} {013043} (\bibinfo {year} {2016})}\BibitemShut
	{NoStop}%
	\bibitem [{\citenamefont {Jaseem}\ \emph {et~al.}(2020)\citenamefont {Jaseem},
		\citenamefont {Hajdu\ifmmode~\check{s}\else \v{s}\fi{}ek}, \citenamefont
		{Solanki}, \citenamefont {Kwek}, \citenamefont {Fazio},\ and\ \citenamefont
		{Vinjanampathy}}]{Jaseem_2020}%
	\BibitemOpen
	\bibfield  {author} {\bibinfo {author} {\bibfnamefont {N.}~\bibnamefont
			{Jaseem}}, \bibinfo {author} {\bibfnamefont {M.}~\bibnamefont
			{Hajdu\ifmmode~\check{s}\else \v{s}\fi{}ek}}, \bibinfo {author}
		{\bibfnamefont {P.}~\bibnamefont {Solanki}}, \bibinfo {author} {\bibfnamefont
			{L.-C.}\ \bibnamefont {Kwek}}, \bibinfo {author} {\bibfnamefont
			{R.}~\bibnamefont {Fazio}},\ and\ \bibinfo {author} {\bibfnamefont
			{S.}~\bibnamefont {Vinjanampathy}},\ }\bibfield  {title} {\bibinfo {title}
		{Generalized measure of quantum synchronization},\ }\href
	{https://doi.org/10.1103/PhysRevResearch.2.043287} {\bibfield  {journal}
		{\bibinfo  {journal} {Phys. Rev. Res.}\ }\textbf {\bibinfo {volume} {2}},\
		\bibinfo {pages} {043287} (\bibinfo {year} {2020})}\BibitemShut {NoStop}%
	\bibitem [{\citenamefont {Barak}\ and\ \citenamefont
		{Ben-Aryeh}(2005)}]{phase_dist_Barak}%
	\BibitemOpen
	\bibfield  {author} {\bibinfo {author} {\bibfnamefont {R.}~\bibnamefont
			{Barak}}\ and\ \bibinfo {author} {\bibfnamefont {Y.}~\bibnamefont
			{Ben-Aryeh}},\ }\bibfield  {title} {\bibinfo {title} {Non-orthogonal positive
			operator valued measure phase distributions of one- and two-mode
			electromagnetic fields},\ }\href {https://doi.org/10.1088/1464-4266/7/5/001}
	{\bibfield  {journal} {\bibinfo  {journal} {Journal of Optics B: Quantum and
				Semiclassical Optics}\ }\textbf {\bibinfo {volume} {7}},\ \bibinfo {pages}
		{123} (\bibinfo {year} {2005})}\BibitemShut {NoStop}%
	\bibitem [{\citenamefont {Susskind}\ and\ \citenamefont
		{Glogower}(1964)}]{PhysicsPhysiqueFizika.1.49}%
	\BibitemOpen
	\bibfield  {author} {\bibinfo {author} {\bibfnamefont {L.}~\bibnamefont
			{Susskind}}\ and\ \bibinfo {author} {\bibfnamefont {J.}~\bibnamefont
			{Glogower}},\ }\bibfield  {title} {\bibinfo {title} {Quantum mechanical phase
			and time operator},\ }\href
	{https://doi.org/10.1103/PhysicsPhysiqueFizika.1.49} {\bibfield  {journal}
		{\bibinfo  {journal} {Physics Physique Fizika}\ }\textbf {\bibinfo {volume}
			{1}},\ \bibinfo {pages} {49} (\bibinfo {year} {1964})}\BibitemShut {NoStop}%
	\bibitem [{Sup()}]{SuppMat}%
	\BibitemOpen
	\href@noop {} {}\bibinfo {note} {See Supplemental Material at [URL] for
		videos of time evolutions shown in Figs.~9 and 10.}\BibitemShut {Stop}%
	\bibitem [{\citenamefont {Weis}\ \emph {et~al.}(2023)\citenamefont {Weis},
		\citenamefont {Fruchart}, \citenamefont {Hanai}, \citenamefont {Kawagoe},
		\citenamefont {Littlewood},\ and\ \citenamefont {Vitelli}}]{weis2023}%
	\BibitemOpen
	\bibfield  {author} {\bibinfo {author} {\bibfnamefont {C.}~\bibnamefont
			{Weis}}, \bibinfo {author} {\bibfnamefont {M.}~\bibnamefont {Fruchart}},
		\bibinfo {author} {\bibfnamefont {R.}~\bibnamefont {Hanai}}, \bibinfo
		{author} {\bibfnamefont {K.}~\bibnamefont {Kawagoe}}, \bibinfo {author}
		{\bibfnamefont {P.~B.}\ \bibnamefont {Littlewood}},\ and\ \bibinfo {author}
		{\bibfnamefont {V.}~\bibnamefont {Vitelli}},\ }\href
	{https://arxiv.org/abs/2207.11667} {\bibinfo {title} {Exceptional points in
			nonlinear and stochastic dynamics}} (\bibinfo {year} {2023}),\ \Eprint
	{https://arxiv.org/abs/2207.11667} {arXiv:2207.11667} \BibitemShut {NoStop}%
	\bibitem [{\citenamefont {Lambert}\ \emph {et~al.}(2024)\citenamefont
		{Lambert}, \citenamefont {Giguère}, \citenamefont {Menczel}, \citenamefont
		{Li}, \citenamefont {Hopf}, \citenamefont {Suárez}, \citenamefont {Gali},
		\citenamefont {Lishman}, \citenamefont {Gadhvi}, \citenamefont {Agarwal},
		\citenamefont {Galicia}, \citenamefont {Shammah}, \citenamefont {Nation},
		\citenamefont {Johansson}, \citenamefont {Ahmed}, \citenamefont {Cross},
		\citenamefont {Pitchford},\ and\ \citenamefont {Nori}}]{QuTiP}%
	\BibitemOpen
	\bibfield  {author} {\bibinfo {author} {\bibfnamefont {N.}~\bibnamefont
			{Lambert}}, \bibinfo {author} {\bibfnamefont {E.}~\bibnamefont {Giguère}},
		\bibinfo {author} {\bibfnamefont {P.}~\bibnamefont {Menczel}}, \bibinfo
		{author} {\bibfnamefont {B.}~\bibnamefont {Li}}, \bibinfo {author}
		{\bibfnamefont {P.}~\bibnamefont {Hopf}}, \bibinfo {author} {\bibfnamefont
			{G.}~\bibnamefont {Suárez}}, \bibinfo {author} {\bibfnamefont
			{M.}~\bibnamefont {Gali}}, \bibinfo {author} {\bibfnamefont {J.}~\bibnamefont
			{Lishman}}, \bibinfo {author} {\bibfnamefont {R.}~\bibnamefont {Gadhvi}},
		\bibinfo {author} {\bibfnamefont {R.}~\bibnamefont {Agarwal}}, \bibinfo
		{author} {\bibfnamefont {A.}~\bibnamefont {Galicia}}, \bibinfo {author}
		{\bibfnamefont {N.}~\bibnamefont {Shammah}}, \bibinfo {author} {\bibfnamefont
			{P.~D.}\ \bibnamefont {Nation}}, \bibinfo {author} {\bibfnamefont {J.~R.}\
			\bibnamefont {Johansson}}, \bibinfo {author} {\bibfnamefont {S.}~\bibnamefont
			{Ahmed}}, \bibinfo {author} {\bibfnamefont {S.}~\bibnamefont {Cross}},
		\bibinfo {author} {\bibfnamefont {A.}~\bibnamefont {Pitchford}},\ and\
		\bibinfo {author} {\bibfnamefont {F.}~\bibnamefont {Nori}},\ }\href
	{https://doi.org/10.48550/arXiv.2412.04705} {\bibinfo {title} {{QuTiP 5: The
				Quantum Toolbox in Python}}} (\bibinfo {year} {2024}),\ \Eprint
	{https://arxiv.org/abs/2412.04705} {arXiv:2412.04705} \BibitemShut {NoStop}%
	\bibitem [{\citenamefont {McKerns}\ \emph {et~al.}(2012)\citenamefont
		{McKerns}, \citenamefont {Strand}, \citenamefont {Sullivan}, \citenamefont
		{Fang},\ and\ \citenamefont {Aivazis}}]{multiprocess}%
	\BibitemOpen
	\bibfield  {author} {\bibinfo {author} {\bibfnamefont {M.~M.}\ \bibnamefont
			{McKerns}}, \bibinfo {author} {\bibfnamefont {L.}~\bibnamefont {Strand}},
		\bibinfo {author} {\bibfnamefont {T.}~\bibnamefont {Sullivan}}, \bibinfo
		{author} {\bibfnamefont {A.}~\bibnamefont {Fang}},\ and\ \bibinfo {author}
		{\bibfnamefont {M.~A.~G.}\ \bibnamefont {Aivazis}},\ }\href
	{https://arxiv.org/abs/1202.1056} {\bibinfo {title} {{Building a Framework
				for Predictive Science}}} (\bibinfo {year} {2012}),\ \Eprint
	{https://arxiv.org/abs/1202.1056} {arXiv:1202.1056} \BibitemShut {NoStop}%
	\bibitem [{\citenamefont {Kehrer}\ and\ \citenamefont {Bruder}(2025)}]{data}%
	\BibitemOpen
	\bibfield  {author} {\bibinfo {author} {\bibfnamefont {T.}~\bibnamefont
			{Kehrer}}\ and\ \bibinfo {author} {\bibfnamefont {C.}~\bibnamefont
			{Bruder}},\ }\href {https://doi.org/10.5281/zenodo.15131083} {\bibinfo
		{title} {{Data for Quantum synchronization blockade induced by nonreciprocal
				coupling}}} (\bibinfo {year} {2025})\BibitemShut {NoStop}%
	\bibitem [{\citenamefont {Wiseman}\ and\ \citenamefont
		{Milburn}(2009)}]{Wiseman2009}%
	\BibitemOpen
	\bibfield  {author} {\bibinfo {author} {\bibfnamefont {H.~M.}\ \bibnamefont
			{Wiseman}}\ and\ \bibinfo {author} {\bibfnamefont {G.~J.}\ \bibnamefont
			{Milburn}},\ }\href {https://doi.org/10.1017/cbo9780511813948} {\emph
		{\bibinfo {title} {{Quantum Measurement and Control}}}}\ (\bibinfo
	{publisher} {Cambridge University Press},\ \bibinfo {year}
	{2009})\BibitemShut {NoStop}%
\end{thebibliography}
\end{document}